%% file: main.tex
\documentclass{article} 
\usepackage{iclr2026_conference,times}
\usepackage{listings}
\usepackage{pgfplots}
\usepackage[most]{tcolorbox}
\pgfplotsset{compat=1.17}
\input{math_commands.tex}

\usepackage{tikz}
\usetikzlibrary{calc} 
\usepackage{hyperref}
\usepackage{bm}
\usepackage{algorithm}      
\usepackage{algpseudocode}  
\usepackage{amssymb}
\usepackage{dsfont}
\usepackage{makecell} 
\usepackage{booktabs}
\usepackage{wrapfig}  
\usepackage{xcolor}   
\usepackage{arydshln}
\usepackage{array}
\usepackage{url}
\usepackage{comment}
\usepackage{graphicx}
\usepackage{lipsum}
\usepackage{subcaption}
\usepackage{enumitem}
\usepackage{multirow}
\usepackage{multicol}

\usepackage{listings}
\definecolor{keywordcolor}{rgb}{0.7, 0.1, 0.1}   
\definecolor{tacticcolor}{rgb}{0.0, 0.1, 0.6}    
\definecolor{commentcolor}{rgb}{0.4, 0.4, 0.4}   
\definecolor{symbolcolor}{rgb}{0.0, 0.1, 0.6}    
\definecolor{sortcolor}{rgb}{0.1, 0.5, 0.1}      
\definecolor{attributecolor}{rgb}{0.7, 0.1, 0.1} 

\newcommand{\ourTool}{\textsc{ProofBridge}}

\title{\ourTool{}: Auto-Formalization of Natural Language Proofs in Lean via Joint Embeddings}

\author{
\textbf{Prithwish Jana}\thanks{Co-corresponding authors. Emails: \texttt{\{pjana7, vganesh\}@gatech.edu}}\,\;$^{1}$\textbf{,}\,\,\,
\textbf{Kaan Kale}$^{1}$\textbf{,}\,\,\,
\textbf{Ahmet Ege Tanriverdi}$^{2}$ \\
\textbf{Cruise Song}$^{1}$\textbf{,}\,\,\,
\textbf{Sriram Vishwanath}$^{1}$\textbf{,}\,\,\, \textbf{Vijay Ganesh}$^{*1}$\\[2mm] 
$^{1}$Georgia Institute of Technology, USA \;\;\;$^{2}$Bogazici University, Türkiye
}

\iclrfinalcopy 
\begin{document}

\maketitle

 \vspace{-1.6mm}
\begin{abstract}
Translating human-written mathematical theorems and proofs from natural language (NL) into formal languages (FLs) like Lean 4 has long been a significant challenge for AI. Most state-of-the-art methods either focus on theorem-only NL-to-FL auto-formalization or on FL proof synthesis from FL theorems. In practice, auto-formalization of both theorem and proof still requires human intervention, as seen in AlphaProof’s silver-medal performance at the 2024 IMO, where problem statements were manually translated before automated proof synthesis.

We present \ourTool{}, a unified framework for automatically translating entire NL theorems and proofs into Lean 4. At its core is a joint embedding model that aligns NL and FL (NL-FL) theorem+proof pairs in a shared semantic space, enabling cross-modal retrieval of semantically relevant FL examples to guide translation.  \ourTool{} integrates retrieval-augmented fine-tuning with iterative proof repair, leveraging Lean’s type checker and semantic equivalence feedback to ensure both syntactic correctness and semantic fidelity. Experiments show substantial improvements in proof auto-formalization over strong baselines (including GPT-5, Gemini-2.5, Kimina-Prover, DeepSeek-Prover), with our retrieval-augmented approach yielding significant gains in semantic correctness (SC, via proving bi-directional equivalence) and type correctness (TC, via type-checking theorem+proof) across pass@k metrics on \textsc{miniF2F-Test-PF}, a dataset we curated. In particular, \ourTool{} improves cross-modal retrieval quality by up to $3.28\times$ Recall@$1$ over all-MiniLM-L6-v2, and achieves +31.14\% SC and +1.64\% TC (pass@32) compared to the baseline Kimina-Prover-RL-1.7B.

\end{abstract}

\section{Introduction}

In mathematics, ensuring the correctness of proofs is a crucial yet inherently difficult task. Traditionally, mathematicians rely on the peer-review process for \textit{proof verification}, yet as proofs grow increasingly complex, even careful human scrutiny can overlook subtle errors. For instance, in 1989, Kapranov and Voevodsky published a proof connecting $\infty$-groupoids and homotopy types, which was later disproven by Carlos Simpson in 1998; more recently, while formalizing his 2023 paper~\citep{tao2023maclaurin} on the Maclaurin-type inequality, Terence Tao discovered a non-trivial bug. To mitigate challenges of verifying complex proofs, proof assistants and formal mathematical languages like Isabelle~\citep{nipkow2002isabelle}, HOL Light~\citep{harrison2009hol}, Metamath~\citep{megill2019metamath}, Lean 4~\citep{moura2021Lean}, Peano~\citep{poesia2023peano}, and Rocq~\citep{rocqprover} have been developed, offering a way to create \textit{computer-verifiable formal proofs}. Such formal language (FL) proofs, defined by strict syntax and symbolic logic, enable reliable automated verification guarantees that resolve the inherent ambiguity of natural language (NL) proofs. However, constructing FL proofs is time-intensive and demands both deep mathematical expertise and detailed knowledge of the language and its libraries. Thus, the process remains challenging even for experienced mathematicians, limiting the wider adoption of formal proof assistants and FL proofs.

\begin{figure}[!t]
    \centering
    \begin{subfigure}[t]{1\textwidth}
        \centering
        \fcolorbox{gray!50}{white}{%
        \includegraphics[width=0.75\linewidth]{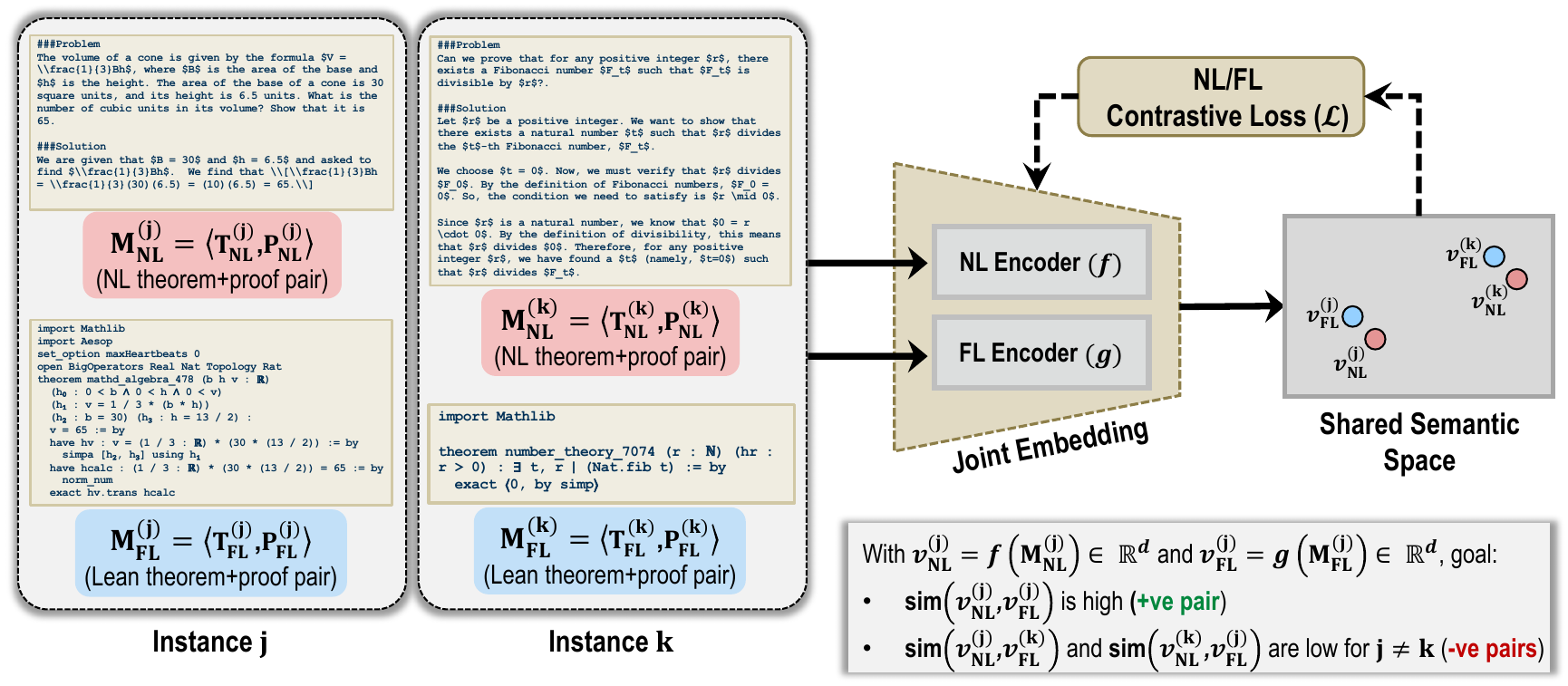}
        }
        \vspace{-1mm}
        \caption{Joint embedding of NL and FL (Lean) theorems and proofs into shared semantic space}
         \label{fig:jtEmbPipeline}
    \end{subfigure}%
    \\\vspace{0mm}
\begin{subfigure}[t]{1\textwidth}
    \centering
    \fcolorbox{gray!50}{white}{%
        \includegraphics[width=0.8\linewidth]{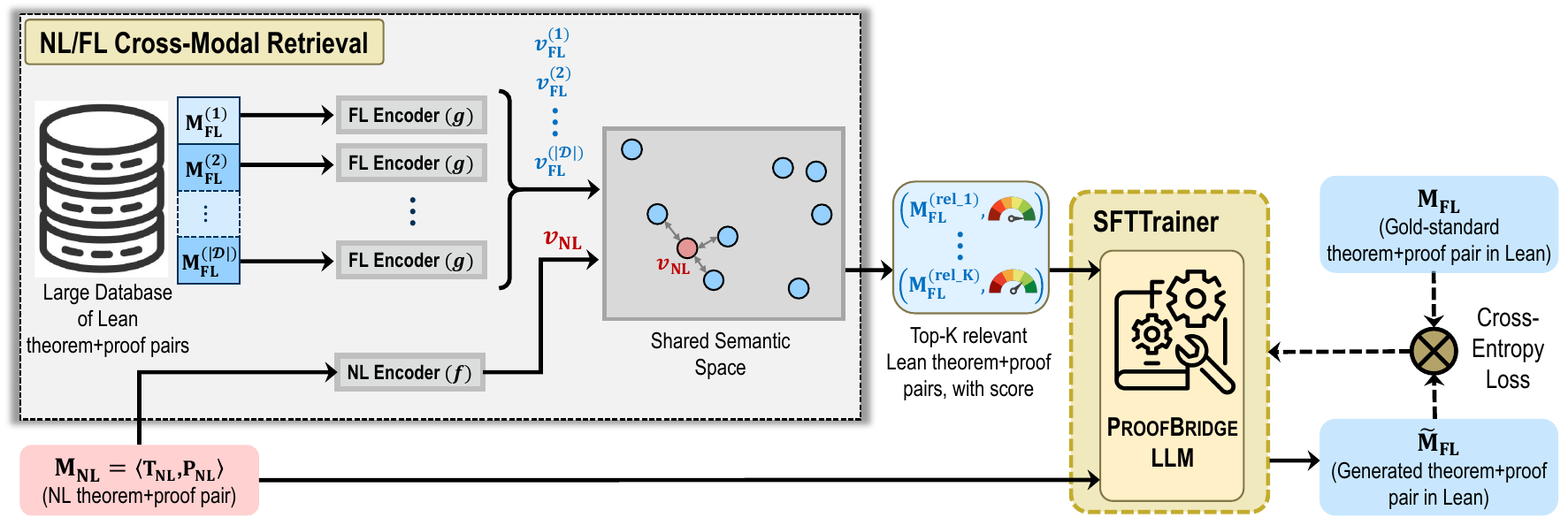}%
    }
    \vspace{-1mm}
    \caption{Retrieval-augmented Supervised Fine-Tuning (SFT) of \ourTool{} with NL/FL cross-modal retrieval}
    \label{fig:sftPipeline}
\end{subfigure}%
\\\vspace{0mm}
\begin{subfigure}[t]{\textwidth}
    \centering
    \fcolorbox{gray!50}{white}{%
        \includegraphics[width=0.8\linewidth]{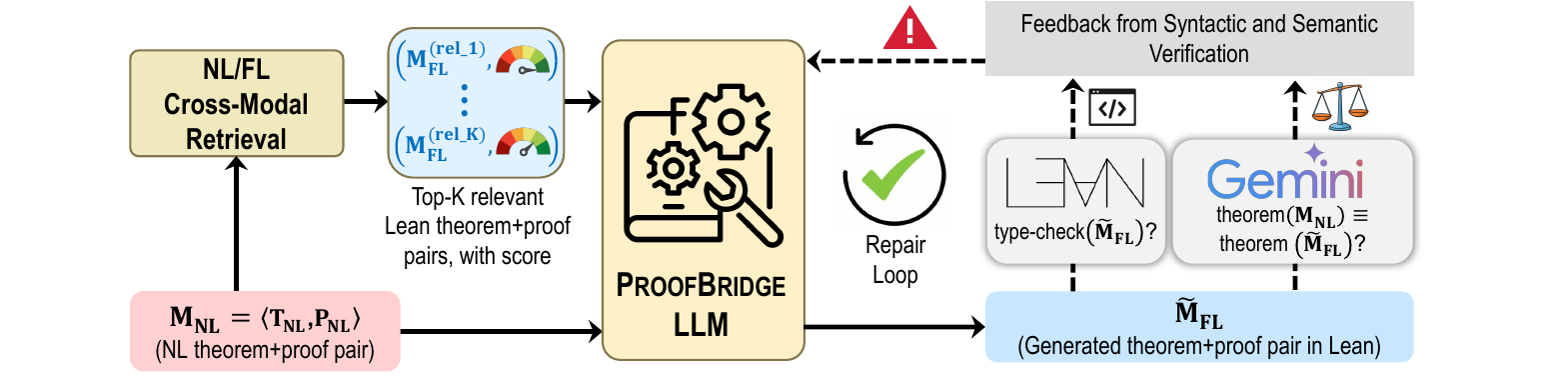}%
    }
    \vspace{-1mm}
    \caption{Inference phase of retrieval-augmented proof auto-formalization with iterative repair}
    \label{fig:inferencePipeline}
\end{subfigure}
\vspace{-2mm}
\caption{\textbf{Pipeline of \ourTool{} for proof auto-formalization.}
We first train a joint embedding model for NL and FL via contrastive learning, enabling cross-modal retrieval of semantically related FL theorem+proof pairs for a given NL input. An LLM is then fine-tuned on NL-to-Lean translations, conditioned on retrieved proofs and relevance scores. At inference, the system retrieves relevant Lean proofs to guide FL theorem+proof generation and applies an iterative repair loop.}
\label{fig:pipeline}
\vspace{-4mm}
\end{figure}
To simplify the task of writing proofs in FL, two key research directions have emerged: \textit{auto-formalization} and \textit{automated formal proof synthesis (AFPS)}. \textit{Auto-formalization} targets NL-to-FL translation, but most prior works~\citep{wang2025kimina,wu2025stepfun,jiang2024multi,gao2024herald} focus only on formalizing theorems (statements), not proofs. In contrast, \textit{AFPS}~\citep{ren2025deepseek,wang2025kimina} aims to generate FL proofs given an FL theorem. Proof auto-formalization is relatively less explored, with Draft-Sketch-Prove~\citep{jiang2022draft} for Isabelle and FormL4~\citep{lu2024process} for Lean serving as notable examples.
In practice, formalizing an entire NL proof requires first performing \textit{theorem-only auto-formalization} to translate the NL theorem into FL, followed by \textit{AFPS} to generate the FL proof. AlphaProof~\citep{deepmind2024ai}, which achieved silver-medal standard in the 2024 International Mathematical Olympiad, followed this two-step process: problems were manually formalized before automated proof synthesis. Thus, in practice, pipelines still require manual theorem formalization prior to proof synthesis, despite advances in theorem-only auto-formalization and AFPS tools. This illustrates the broader challenge that current systems often rely on human intervention to ensure semantic correctness of proof auto-formalization. 


\vspace{-1mm}
Contemporary LLMs face several challenges that limit their effectiveness for proof auto-formalization in Lean 4. \textbf{First}, large-scale datasets pairing NL theorems with Lean 4 proofs are scarce. Most existing resources (Goedel-Pset-v1~\citep{lin2025goedel}, Herald-statements~\citep{gao2024herald}, Lean Workbook~\citep{ying2024lean}, MMA~\citep{jiang2024multi}) cover only theorems, while those with proofs (Herald-proofs, Lean Workbook proofs~\citep{lin2025goedel}, and FormL4~\citep{lu2024process}) are much smaller and do not align with the popular miniF2F~\citep{zheng2021minif2f} benchmark in the same Lean 4 version. Lean versions are not backward compatible, so cross-version evaluation often fails. \textbf{Second}, general-purpose foundation models often fail to satisfy strict syntactic and semantic constraints of specialized FLs~\citep{jha2025rlsf, delorenzo2025abstraction, jana2024neurosymbolic, ugare2024syncode}, such as Lean, and are computationally expensive~\citep{dasgupta2025hitgram}, motivating smaller specialized models. However, most fine-tuned LLMs for Lean target either theorem auto-formalization or proof synthesis. Proof auto-formalization is harder, as it requires both translating the NL theorem and constructing the corresponding FL proof. \textbf{Third}, Lean 4 has an effectively \textit{infinite action space}~\citep{poesia2023peano}, where proofs use complex \textit{tactics} that reuse prior theorems or introduce new variables. Prior work generates FL directly from NL while ignoring semantic relations such as shared tactics and DAG dependencies, often leading LLMs to produce hallucinated or invalid proofs. \textbf{Fourth}, automated evaluation is a major bottleneck. Lean's type-checker verifies the FL proof but cannot ensure semantic equivalence. Existing methods often type-check only the theorem (leaving proofs incomplete using placeholders like \lstinline{sorry}) or rely on proxies such as BLEU, which are unreliable~\citep{lu2024process,ying2024lean,wu2022autoformalization}.

\textbf{Key Insight.} In this paper, we address the task of \textit{proof auto-formalization}, focusing on Lean~4 as the FL, via a combination of a joint embedding model, an LLM, and Lean for verification. It takes as input an NL theorem+proof pair and produces the corresponding FL theorem+proof pair in Lean~4. The key insight behind our approach is to treat proof auto-formalization as \textit{learning from demonstrations}: the LLM is guided not only by the NL proof but also by FL proofs retrieved using an NL/FL joint embedding model that leverages contrastive learning and encodes linear DAG traversals of Lean proofs. Rather than relying on randomly chosen few-shot examples, these retrieved proofs capture far richer \textit{reusable formalization patterns} (tactic choices, DAG structures), providing grounded signals that guide generation toward Lean-verifiable proofs, as illustrated in Figure~\ref{fig:pipeline}.

\noindent\textbf{Contributions:}
\vspace{-1mm}

\noindent{\bf The \ourTool{} Auto-formalization Method and Tool:} We present \ourTool{}, an LLM-based, retrieval-augmented proof auto-formalization framework. At its core is an \textit{NL/FL joint embedding model} that maps semantically equivalent NL and FL theorem+proof pairs to nearby points in a shared space, enabling cross-modal retrieval of related FL proofs. We then fine-tune the SoTA LLM Kimina-Prover-RL-1.7B~\citep{wang2025kimina} to perform NL-to-Lean~4 proof translation, conditioned on the retrieved FL proofs and their relevance scores. During inference, generation is refined with an iterative verifier-guided repair loop combining Lean type-checking with LLM-based bi-directional equivalence proving to ensure syntactic correctness and semantic fidelity. (Section~\ref{sec:proposed})

\vspace{-1mm}
\noindent {\bf \textsc{NuminaMath-Lean-PF} Dataset:} We curate \textsc{NuminaMath-Lean-PF}, a large-scale dataset of 38.9k NL$\leftrightarrow$Lean~4 theorem+proof pairs, specialized for \textit{proof auto-formalization}. Each Lean theorem+proof pair is type-checked and paired with an NL counterpart. Additionally, we release \textsc{miniF2F-Test-PF}, a Lean~v4.15.0 version of miniF2F-Test with 244 instances tailored for proof auto-formalization, enabling a consistent pipeline in the same Lean version. (Section~\ref{sec:train_dataset_curation})

\vspace{-1mm}
\noindent {\bf Extensive Experimental Evaluation:} Compared to the baseline encoder all-MiniLM-L6-v2, \ourTool{}'s cross-modal NL$\to$FL retrieval achieves $3.28\times$ higher Recall Rate@$K$ at $K$=1 and $2.74\times$ MRR, with top-$K$ retrieved embeddings $23\%$ closer and non-retrieved $104\%$ farther.
 We evaluate \ourTool{} against 13 SoTA LLMs, including foundation models (Gemini-2.5, GPT-5-mini) and automated proof synthesis LLMs (DeepSeek-Prover, STP, Leanabell-Prover, Kimina-Prover), using verifier-grounded metrics: \textit{type correctness} (TC) and \textit{semantic correctness} (SC, a new metric based on Lean bidirectional equivalence proofs). Built on Kimina-Prover-RL-1.7B, \ourTool{} achieves +31.14\% SC and +1.64\% TC (pass@32) on \textsc{miniF2F-Test-PF}.
(Section~\ref{sec:expt})
\section{Related Work}
\label{sec:relwork}

Our work lies at the intersection of three key AI-for-Math research areas: automated formal proof synthesis, auto-formalization, and retrieval-augmented learning for mathematical reasoning. We focus on the most relevant approaches and highlight differences from our unified framework.

\textbf{Auto-Formalization.} Auto-formalization translates NL mathematics into FL, but most existing work focuses on theorem formalization rather than proofs. \textbf{\textit{Theorem-only approaches}} include Herald-translator~\citep{gao2024herald}, which extracts FL theorems from Mathlib4 and trains on informal counterparts, and Kimina-Autoformalizer~\citep{wang2025kimina}, which fine-tunes models with expert iteration. These excel at translating theorems but cannot handle proofs. \textbf{\textit{Proof auto-formalization}} has received limited attention. Draft-Sketch-Prove~\citep{jiang2022draft} converts NL proofs into formal sketches in Isabelle with open conjectures, then fills gaps using predefined tactics and tools like Sledgehammer~\citep{paulsson2012three}. FormL4~\citep{lu2024process} trains on GPT-4 informalized Mathlib proofs with process-supervised step-level Lean compilation feedback.

\vspace{-1.2mm}
\emph{Key Differences:} Our approach is the first to jointly learn representations for NL and FL theorem+proof pairs, enabling cross-modal retrieval to guide formalization. Unlike prior work, we leverage semantic relationships of NL and FL proofs for contextual guidance.

\textbf{Automated Formal Proof Synthesis (AFPS).} In this setting, given an FL theorem, the goal is to generate its FL proof. \textbf{\textit{Next-Tactic Prediction (NTP)}} methods predict each proof step from the current proof state. Examples include GPT-f~\citep{polu2020generative} for Metamath, LIsa~\citep{jiang2021lisa} for Isabelle, and PACT~\citep{han2022proof} for Lean. They use tree search over proof states, prioritizing tactics by cumulative probability. While NTP ensures stepwise correctness via interactive theorem-prover verification, it suffers from long-horizon dependencies and repeated verification overhead. \textbf{\textit{Whole-Proof Generation (WPG)}} methods generate complete FL proofs in a single pass, offering computational efficiency but risking cascading errors. Recent advances include DeepSeek-Prover-v1~\citep{xin2024deepseek}, which combines SFT with expert iteration, and TheoremLlama~\citep{wang2024theoremllama}, which improves in-context learning via curriculum-based training. DeepSeek-Prover-v2~\citep{ren2025deepseek} integrates NL reasoning with FL proof generation, while Kimina-Prover~\citep{wang2025kimina} applies reinforcement learning with compilation-based rewards~\citep{jana2024cotran}.

\vspace{-1mm}
\emph{Key Differences:} Unlike AFPS approaches that assume FL theorems as input, our work addresses the more challenging task of generating theorem+proof pairs in FL from an NL input.

\textbf{Retrieval-Augmented Learning for Mathematics.} Recent work has explored retrieval-augmented mathematical reasoning, though not specifically for auto-formalization. TLAPS~\citep{zhou2025retrieval} retrieves verified proofs to aid proof generation, while COPRA~\citep{thakur2023language} and REAL-Prover~\citep{shen2025real} retrieve lemmas or theorems to guide proof search. These methods rely on plain-text encoding. LeanSearch~\citep{gao2024semantic}, HERALD~\citep{gao2024herald}, and RAutoformalizer~\citep{liu2025rethinking} also use plain-text encoders for FL theorem retrieval; however, as shown in Section~\ref{sec:exptresults}, this does not extend to the more demanding task of FL theorem+proof pair retrieval.

\vspace{-1mm}
\emph{Key Differences:} Our joint embedding enables NL/FL cross-modal retrieval of theorem+proof pairs and encodes the DAG structure of Lean proofs, unlike plain-text encoders. Capturing proof-structure semantics is essential for proof auto-formalization and is not addressed by existing tool-chains.

\textbf{Positioning our contributions.} The combination of joint embedding, retrieval augmentation, and unified translation distinguishes \ourTool{} from prior work: \textit{(a) Unified Proof Auto-Formalization:} We address complete translation (theorem + proof) rather than treating theorem formalization and proof synthesis separately. \textit{(b) Joint Semantic Embedding:} Our contrastive learning framework for aligning NL and FL proofs is novel, enabling effective cross-modal retrieval. \textit{(c) Retrieval-Augmented Translation:} We are the first to apply retrieval-augmented fine-tuning and generation to auto-formalization, leveraging semantic relationships of FL proofs to guide translation. \textit{(d) Rigorous Evaluation:} We introduce systematic metrics for proof auto-formalization, covering type checking and semantic correctness via bi-directional equivalence proofs instead of proxy measures.

\section{Preliminaries: Tactic-style Proofs in Lean}
\label{sec:prelim}

\begin{wrapfigure}[15]{r}[0pt]{0.38\textwidth} 
\vspace{-4mm}
  \centering
    \includegraphics[width=\linewidth]{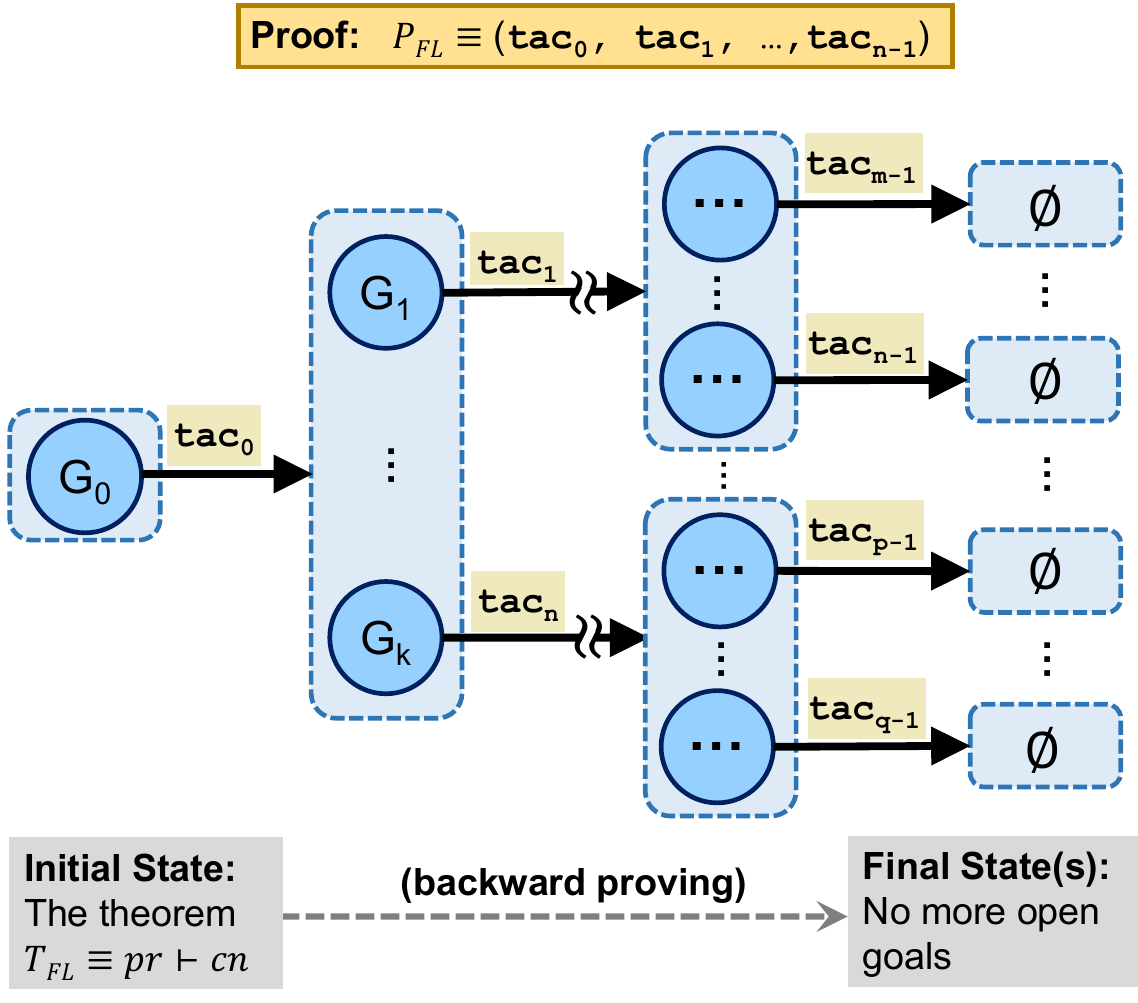}%
\caption{\textbf{Tactic-style Proof.} A Lean proof represented as a DAG of tactics.}
  \label{fig:tacstyleproofs}
\end{wrapfigure}

Lean~\citep{moura2021Lean} is a functional programming language and interactive theorem prover that is based on the propositions-as-types principle, where proving a proposition is equivalent to constructing a term of the corresponding type. Rather than building these terms manually, users write proofs in a tactic language, which provides high-level steps to guide term construction. Lean~4 (henceforth Lean) represents tactic-style proofs as directed acyclic graphs (DAGs) of \textit{proof states} and \textit{tactics}, automatically generating the corresponding proof term in the background. The kernel then verifies the term, ensuring correctness by enforcing the axiomatic foundations of Lean's logic, the Calculus of Inductive Constructions. This combination of a formal system and a small, trusted kernel provides strong confidence in the validity of proofs. In the DAG (Figure~\ref{fig:tacstyleproofs}) of a Lean proof:

\setlength{\columnsep}{9pt}
\setlength{\fboxsep}{0pt} 

\begin{itemize}[left=0em, topsep=0em, itemsep=0em, parsep=0em]
    \item Each \textbf{proof state}  $S_i\equiv\left[G_1,\cdots,G_n\right]$ consists of a sequence of zero or more \textit{open goals}. Initial state $S_0$ has one goal, the theorem $T_{\mathit{FL}}\equiv\mathit{pr}\vdash\mathit{cn}$ itself. Leaf-level states have no open goal.
    \item Each \textbf{open goal} $G_i\equiv\mathit{pr}_i\vdash\mathit{cn}_i$ of a proof state represents a proposition $\textit{cn}_i$ that needs to be proven, given a set of premises $\textit{pr}_i$. 
    \item Each \textbf{tactic} $\textit{tac}_i$ represents a proof step. It is a high-level command (rooted in metaprogramming) applied to an open goal $G_i$, producing a new proof state. If the resulting proof state has no open goal, it directly resolves the current goal. A parent goal is resolved once all subgoals are resolved.
\end{itemize}

\noindent Tactic-style proof synthesis in Lean follows a \textit{sequential decision process}. Lean provides an interactive REPL~\citep{lean4repl} that applies tactics step by step to manipulate proof states. An FL proof is a sequence of tactics, and at each step, the REPL updates the proof state if the tactic is valid or returns an error identifying the faulty one. Each tactic advances the proof by breaking the current goal into simpler subgoals, similar to the `\textit{suffices to show}' construct.


\textbf{Proof Auto-formalization as a Learning Problem.} 
Given an NL theorem+proof pair $M_{\text{NL}} = \langle T_{\text{NL}}, P_{\text{NL}} \rangle$, the goal is to learn a function $f \colon M_{\text{NL}} \mapsto M_{\text{FL}}$ that produces a corresponding Lean theorem+proof pair $M_{\text{FL}} = \langle T_{\text{FL}}, P_{\text{FL}} \rangle$. Here, $T_{\text{FL}} \equiv \mathit{pr} \vdash \mathit{cn}$ denotes the formal theorem, and $P_{\text{FL}} \equiv (\texttt{tac}_0, \dots, \texttt{tac}_{n-1})$ is the proof as a sequence of tactics. The generated pair must satisfy:

\vspace{-1mm}
\begin{enumerate}[label=(\alph*),left=0em, topsep=0em, itemsep=0em, parsep=0em]
    \item \textit{Type correctness:} $M_{\text{FL}}=\langle T_{\text{FL}}, P_{\text{FL}}\rangle$ passes Lean type-checking, ensuring that $P_{\text{FL}}$ proves $T_{\text{FL}}$ with no open goals in the DAG.
    \item \textit{Semantic correctness:} FL theorem is semantically equivalent to the NL one, i.e., $T_{\text{FL}} \equiv T_{\text{NL}}$.
\end{enumerate}


\section{Our Approach and Tool Architecture}
\label{sec:proposed}


\subsection{Joint Embedding of NL and Lean Proofs for Cross-Modal Retrieval}
A core component of our framework is the \textit{joint embedding model}, which learns to represent NL theorem+proof pairs and their FL (Lean) counterparts in a shared semantic space. The goal is to align these modalities so that cross-modal retrieval between NL and FL becomes possible. Let $M_{\text{FL}}^{(i)} = \langle T_{\text{FL}}^{(i)}, P_{\text{FL}}^{(i)} \rangle$ denote an FL theorem+proof pair, and let $\mathcal{D} = \{ M_{\text{FL}}^{(i)} \}_{i=1}^{N}$ be a large database of such pairs. Given an NL theorem+proof pair $M_{\text{NL}}$, the model retrieves a subset $\mathcal{R}(M_{\text{NL}}, \mathcal{D}) \subset\mathcal{D}$ of size $K \ll |\mathcal{D}|$ that serves as in-context demonstrations to guide downstream auto-formalization.

During training, $M^{(i)}_{\text{NL}}$ and $M^{(i)}_{\text{FL}}$ are encoded into vectors by modality-specific encoders. Each encoder is initialized with a pre-trained model, and a subset of parameters is subsequently fine-tuned. The NL encoder $f$ maps $M^{(i)}_{\text{NL}}$ to a vector $v^{(i)}_{\text{NL}} = f(M^{(i)}_{\text{NL}}, \theta_f \Vert \phi_f) \in \mathbb{R}^d$, and the FL encoder $g$ maps $M^{(i)}_{\text{FL}}$ to $v^{(i)}_{\text{FL}} = g(M^{(i)}_{\text{FL}}, \theta_g \Vert \phi_g) \in \mathbb{R}^d$, where $\theta$ and $\phi$ denote frozen and trainable parameters, respectively, and $d$ is the shared embedding dimension. The details of each encoder are as follows:


\begin{itemize}[left=0em, topsep=0em, itemsep=0em, parsep=0em]
    \item \textbf{NL encoder} $\bm{f(M^{(i)}_{\text{NL}}, \theta_f \Vert \phi_f)}$\textbf{:} To encode $M^{(i)}_{\text{NL}}$, we use \texttt{all-MiniLM-L6-v2}~\citep{sentence_transformers}, a lightweight model (22.7M parameters) that effectively captures semantic similarity in NL. It encodes $M^{(i)}_{\text{NL}}$ into $384$-dimensional embeddings, thereby projected into the joint embedding space of dimension $d=512$ via a linear layer included in the trainable set $\phi_{\text{f}}$.

    \item \textbf{FL encoder $\bm{g(M^{(i)}_{\text{FL}}, \theta_g \Vert \phi_g)}$:} Given $M_{\text{FL}}^{(i)} = \langle T_{\text{FL}}^{(i)}, P_{\text{FL}}^{(i)} \rangle$, we first extract the linearized DAG traversal of tactics from $P_{\text{FL}}^{(i)}$ using Lean REPL~\citep{lean4repl}. This traversal is represented as an ordered sequence of proof-state transformations induced by successive tactic applications: $S_0 \xrightarrow{\texttt{tac}_0} S_1 \xrightarrow{\texttt{tac}_1} \cdots \xrightarrow{\texttt{tac}_{H-1}} S_H$, where $S_0 \equiv T_{\text{FL}}^{(i)}$, each $S_h \equiv [G_1, \dots, G_l]$ denotes a proof state consisting of zero or more open goals, and $\texttt{tac}_{h-1}$ is the tactic applied at step $h$. This sequence captures the entire proof as an ordered series of state transformations. To create embeddings for the full proof, we first encode each state $S_h$ using LeanDojo's \texttt{ByT5} proof-state encoder~\citep{yang2023leandojo} (218M parameters), producing embeddings of size $1{,}472$ per state. We then obtain a single embedding for the entire proof via mean-pooling, which is subsequently projected into a shared semantic space of dimension $d = 512$ using a linear layer included in the trainable parameters $\phi_{\text{g}}$. The intuition behind this approach is to ensure that semantically similar proofs (those with similar DAG structures of proof states and tactics) produce similar embeddings.
\end{itemize}

\vspace{-1mm}
\textbf{Contrastive Learning.} To enable cross-modal retrieval between NL and FL representations, it is essential to align the two modalities in the embedding space. Specifically, for each positive pair $\big(M^{(i)}_{\text{NL}}, M^{(i)}_{\text{FL}}\big)$, we aim for their embeddings $\big(v^{(i)}_{\text{NL}}, v^{(i)}_{\text{FL}}\big)$ to exhibit high cosine similarity, while embeddings of mismatched pairs are pushed apart. Denoting $\ell_2$-normalization by $\widehat{v}=v/\|v\|_2$ and defining the cosine similarity between two embeddings $u$ and $w$ as $\left[ \widehat{u},\widehat{w}\right]$, we adopt the following symmetric contrastive loss for a mini-batch $\mathcal{B} = \big\{\big(M^{(i)}_{\text{NL}}, M^{(i)}_{\text{FL}}\big)\big\}_{i=1}^{n}$ of NL and FL pairs:

\vspace{-4mm}
\begin{equation}
\mathcal{L}(\mathcal{B})
=
-\frac{1}{2n}\sum_{i=1}^{n}
\Bigg[
\log\left(\frac{
\exp\!\big(\big[ \widehat{v}^{(i)}_{\text{NL}},\widehat{v}^{(i)}_{\text{FL}}\big]/\tau\big)
}{
\sum_{j=1}^{n}\exp\!\big(\big[ \widehat{v}^{(i)}_{\text{NL}},\widehat{v}^{(j)}_{\text{FL}}\big]/\tau\big)
}\right)
+
\log\left(\frac{
\exp\!\big(\big[ \widehat{v}^{(i)}_{\text{FL}},\widehat{v}^{(i)}_{\text{NL}}\big]/\tau\big)
}{
\sum_{j=1}^{n}\exp\!\big(\big[ \widehat{v}^{(i)}_{\text{FL}},\widehat{v}^{(j)}_{\text{NL}}\big]/\tau\big)
}\right)
\Bigg]
\label{equation:contrastive-objective}
\end{equation}
\vspace{-2mm}

where $\tau>0$ is a temperature hyperparameter. This loss encourages each NL embedding to be closest to its corresponding FL embedding, and vice versa, using other in-batch embeddings as negatives. The negatives are sampled randomly for each mini-batch.

\vspace{-1.5mm}
\textbf{NL/FL Cross-Modal Retrieval.} We precompute the normalized embeddings 
$\big\{\widehat{v}^{(i)}_{\mathrm{NL}}\big\}_{i=1}^{|\mathcal{D}|}$  and $\big\{\widehat{v}^{(j)}_{\mathrm{FL}}\big\}_{j=1}^{|\mathcal{D}|}$ for all NL and FL theorem+proof pairs respectively in our database $\mathcal{D}$, which enables efficient cross-modal retrieval. 
Given a query theorem+proof pair in either source modality (NL or FL), we encode it into the shared semantic space (yielding $\widehat{q}_{\mathrm{NL}}$ or $\widehat{q}_{\mathrm{FL}}$) and compute cosine similarities with all items in the target modality, producing the set
$\big\{\big[\widehat{q}_{\mathrm{NL}}, \widehat{v}^{(j)}_{\mathrm{FL}}\big]\big\}_{j=1}^{|\mathcal{D}|}$ or $\big\{\big[\widehat{q}_{\mathrm{FL}}, \widehat{v}^{(i)}_{\mathrm{NL}}\big]\big\}_{i=1}^{|\mathcal{D}|}$, 
depending on the retrieval direction. The top-$K$ nearest neighbors from these sets are then selected as demonstrations, reflecting similar proof structures, patterns, and mathematical domains.


\subsection{Retrieval-Augmented Fine-Tuning for Proof Auto-Formalization}

We fine-tune an LLM to translate NL theorem+proof pairs into FL (Lean), conditioned on retrieved FL demonstrations that provide rich contextual knowledge. For each training instance, we construct a prompt containing (a) input NL theorem+proof pair $M_{\text{NL}}$ and (b) top-$K$ retrieved FL theorem+proof pairs: $\mathcal{R}(M_{\text{NL}}, \mathcal{D})=\{ M^{(k)}_{\text{FL}} \}_{k=1}^{K}$ with relevance scores $\{r^{(k)}\}_{k=1}^{K}$. The retrieved examples demonstrate how similar mathematical concepts and proof strategies are formalized in Lean. We include relevance scores to help the model weight the importance of each retrieved example.



\textbf{Training Objective.} We fine-tune Kimina-Prover-RL-1.7B~\citep{wang2025kimina} using supervised learning on our \textsc{NuminaMath-Lean-PF} dataset (details in Section~\ref{sec:train_dataset_curation}). The model is trained to generate an FL theorem+proof pair $\widetilde{M}_{\text{FL}}$ given the input context. This retrieval-augmented approach allows the LLM to learn from similar formalization patterns rather than generating formal theorems in isolation. As illustrated in Figure~\ref{fig:sftPipeline}, we use the standard auto-regressive language modeling loss:

\vspace{-5mm}
\begin{equation}
\mathcal{L}_{\text{CE}} = -\frac{1}{|\mathcal{T}|} \sum_{t=1}^{|\mathcal{T}|} \log P_\theta\left( \tau_t \mid \tau_{<t}, \mathcal{C} \right)
\end{equation}
\vspace{-4mm}

where $\mathcal{T} = \widetilde{M}_{\text{FL}}$ is the generated formalization tokenized as sequence $(\tau_1, \ldots, \tau_{|\mathcal{T}|})$, $\mathcal{C}$ represents the input context (NL theorem+proof and retrieved FL examples), and $\theta$ are the LLM parameters. This corresponds to the cross-entropy loss between $\widetilde{M}_{\text{FL}}$ and the gold-standard formalization $M_{\text{FL}}$.

\subsection{Iterative Proof Repair with Verifier Feedback}

During inference, we perform retrieval-augmented proof auto-formalization with the fine-tuned LLM (Figure~\ref{fig:inferencePipeline}). However, LLM being a stochastic model may still generate FL theorem+proof pair that contain syntactic errors or semantic misalignments with the input NL theorem+proof. To address, we implement an iterative repair mechanism that combines Lean's type checker with semantic equivalence verification. For an input NL theorem+proof pair $M_{\text{NL}}=\langle T_{\text{NL}}, P_{\text{NL}}\rangle$ the LLM generates an FL counterpart $\widetilde{M}_{\text{FL}}=\langle \widetilde{T}_{\text{FL}}, \widetilde{P}_{\text{FL}}\rangle$, on which we perform two types of verification:

\algrenewcommand\alglinenumber[1]{\scriptsize #1}
\begin{wrapfigure}[11]{r}{0.5\textwidth} 
\begingroup
\scriptsize 
\vspace{-9mm}
\begin{minipage}{\linewidth}
\begin{algorithm}[H]
\caption{Iterative Proof Repair}
\begin{algorithmic}[1]
\scriptsize
\State \textbf{Input:} NL theorem+proof pair 
$M_{\text{NL}}=\langle T_{\text{NL}}, P_{\text{NL}}\rangle$, 
\Statex \phantom{\textbf{Input:}} Initial FL theorem+proof pair 
$\widetilde{M}^{(0)}_{\text{FL}}=\langle \widetilde{T}^{(0)}_{\text{FL}}, \widetilde{P}^{(0)}_{\text{FL}}\rangle$
\State \textbf{Output:} Verified FL theorem+proof pair or FAILURE
\For{$i = 0$ to $R_{\max} - 1$}
    \State $\text{syntaxOK} \leftarrow \text{LeanTypeCheck}\big(\widetilde{M}^{(i)}_{\text{FL}}\big)$
    \State $\text{semanticsOK} \leftarrow \text{SemanticEquivalence}\big(T_{\text{NL}}, \widetilde{T}^{(i)}_{\text{FL}}\big)$
    \State \textbf{if} $\text{syntaxOK} \land \text{semanticsOK}$ \textbf{then return} $\widetilde{M}^{(i)}_{\text{FL}}$
    \State $\text{feedback} \leftarrow \text{GenerateFeedback}(\text{syntaxOK}, \text{semanticsOK})$
    \State $\widetilde{M}^{(i+1)}_{\text{FL}} \leftarrow \text{LLMRepair}(\text{feedback})$
\EndFor
\State \textbf{return} FAILURE
\end{algorithmic}
\label{algo:proof-repair}
\end{algorithm}
\end{minipage}
\endgroup
\end{wrapfigure}

\vspace{-1mm}
\begin{enumerate}[left=0em, topsep=0em, itemsep=0em, parsep=0em]
    \item \textbf{Syntactic Verification:} We compile $\widetilde{M}_{\text{FL}}$ using Lean's type checker. If compilation fails, we extract the specific error message and location from Lean's diagnostic output.
    
    \item \textbf{Semantic Verification:} We assess whether the generated theorem $\widetilde{T}_{\text{FL}}$ accurately represents the original NL theorem $T_{\text{NL}}$ using an LLM-based equivalence judge.
\end{enumerate}

\textbf{Repair Process.} When either syntactic or semantic verification fails, we initiate an iterative repair process. The procedure terminates once both checks succeed or the maximum number of repair attempts ($R_{\max}=5$) is reached. This bounded, iterative strategy improves the reliability of proof auto-formalization by catching and correcting common errors while maintaining computational efficiency. The overall process is described in Algorithm~\ref{algo:proof-repair}.

\section{Experimental Evaluation}
\label{sec:expt}

\subsection{Datasets and Preparation: \textsc{NuminaMath-Lean-PF} and \textsc{miniF2F-Test-PF}}
\label{sec:train_dataset_curation}

For training, we construct \textsc{NuminaMath-Lean-PF} from NuminaMath-LEAN~\citep{wang2025kimina}, containing 104,155 competition-level problems in algebra, geometry, number theory, combinatorics, and calculus. Each instance pairs an NL theorem $T_{\text{NL}}$ with a human-written Lean v4.15.0 theorem $T_{\text{FL}}$; 38,951 include FL proofs $P_{\text{FL}}$ (30\% human-written, rest by KiminaProver), forming $\{T_{\text{NL}}, \langle T_{\text{FL}}, P_{\text{FL}}\rangle\}$. Next, we prepare \textsc{NuminaMath-Lean-PF} via the following steps:

\vspace{-1mm}
\textbf{Formal Verification and Repair.} Each $\langle T_{\text{FL}}, P_{\text{FL}}\rangle$ is type-checked using Lean REPL~\citep{lean4repl}. About 6\% (2,337) failed due to syntax, library mismatches, or incomplete proofs. These were automatically repaired via Gemini-2.5-Pro: error messages and locations are extracted from Lean, used to prompt the LLM for corrections, and re-verified iteratively up to five times.


\vspace{-1.5mm}
\textbf{NL Proof Generation.} As NuminaMath-LEAN provides only theorems, we generate NL proofs in two stages. First, \textit{solution sketch retrieval} searches for $T_{\text{NL}}$ in NuminaMath 1.5~\citep{numina_math_datasets} (896k problem-solution pairs), retrieving NL solution sketches (median 79 words) for 25,792 instances (66\%). Next, \textit{FL-to-NL informalization} uses Gemini-2.5-Pro to translate $P_{\text{FL}}$ into NL proofs, using available sketches, yielding 38,951 NL-FL theorem+proof pairs \(\{\langle T_{\text{NL}},P_{\text{NL}}\rangle,\langle T_{\text{FL}},P_{\text{FL}}\rangle\}\).


For validation, we curate \textsc{miniF2F-Test-PF} by combining two versions of miniF2F-test~\citep{zheng2021minif2f}, a widely-used auto-formalization benchmark. It contains 244 Olympiad-level problems from the AIME, AMC, IMO, and undergraduate courses in algebra, number theory, and inequalities. We use the Lean v4.15.0 version~\citep{AIMO_minif2f_test_2025} and add missing NL proofs from~\cite{yangky11_miniF2F_lean4_2025}.

\subsection{Evaluation Metrics}

\textbf{Metrics for NL/FL Cross-Modal Retrieval.} 
We evaluate cross-modal alignment of our joint embedding model in two directions. NL $\to$ FL measures retrieval of FL theorem+proof pairs given an NL input, which is relevant for proof auto-formalization, while FL $\to$ NL assesses the reverse. For a test pair $\big(M_{\text{NL}}, M_{\text{FL}}\big)$, a retrieval in the NL $\to$ FL direction is deemed successful if the model retrieves the FL counterpart $M_{\text{FL}}$ given $M_{\text{NL}}$, and unsuccessful otherwise; the FL $\to$ NL direction is evaluated analogously. We assess retrieval performance using five metrics. \textbf{\textit{(i) Recall Rate @}} $\bm{K}$ measures the percentage of queries for which the query's cross-modal counterpart appears among the top-$K$ retrieved results. We report $K = 1, 5, 10, 20, 50$. \textbf{\textit{(ii) Mean Reciprocal Rank (MRR)}} is the average reciprocal rank of the retrieved cross-modal counterpart for each query, $\operatorname{MRR} = \frac{1}{N} \sum_{q=1}^{N} \frac{1}{\operatorname{rank}_q}$, indicating how highly it is ranked. \textbf{\textit{(iii) Cosine Similarity of Top-$\bm{K}$ Retrieved}} measures the cosine similarity between the query embedding and those of the top-$K$ retrieved instances. For each query, we sort these scores in ascending order and record three statistics: median ({\small \texttt{\textbf{M}}}), $25^\text{th}$ percentile ({\small \texttt{\textbf{Q1}}}), and $75^\text{th}$ percentile ({\small \texttt{\textbf{Q3}}}), and report their average over the test set. \textbf{\textit{(iv) Cosine Similarity of Non-Retrieved}} applies the same procedure to all non-retrieved instances and reports the median ({\small \texttt{\textbf{M}}}), $25^\text{th}$ percentile ({\small \texttt{\textbf{Q1}}}), and $75^\text{th}$ percentile ({\small \texttt{\textbf{Q3}}}) averaged over the test set. \textbf{\textit{(v) mean Median Gap (mMG)}} measures the difference between the median ({\small \texttt{\textbf{M}}}) cosine similarity of top-$K$ retrieved and that of non-retrieved instances, averaged over $K = 1,5,10,20,50$.

\textbf{Metrics for Proof Auto-Formalization.} 
Given $M_{\text{NL}} = \langle T_{\text{NL}}, P_{\text{NL}}\rangle$, a proof auto-formalizer generates an FL version $\smash{\widetilde{M}_{\text{FL}} = \langle \widetilde{T}_{\text{FL}}, \widetilde{P}_{\text{FL}}\rangle}$. We evaluate this using two metrics: \textbf{\textit{(i) Type Correctness (TC)}} checks if $\smash{\widetilde{M}_{\text{FL}}}$ passes Lean's type-checker, i.e., $\smash{\widetilde{P}_{\text{FL}}}$ proves $\smash{\widetilde{T}_{\text{FL}}}$ without using keyword \lstinline{sorry}. \textbf{\textit{(ii)
Semantic Correctness (SC)}} is measured only for type-correct outputs and checks whether $\smash{\widetilde{T}_{\text{FL}}}$ is definitionally equal to $T_{\text{FL}}$ (we admit some propositional equalities) by prompting Gemini-2.5-Pro up to five times to produce a Lean proof of the bi-directional equivalence $\smash{\widetilde{T}_{\text{FL}} \leftrightarrow T_{\text{FL}}}$ using a restricted set of tactics, e.g., \lstinline{rfl}, \lstinline{simp}, \lstinline{ring}, etc. (see Appendix~\ref{appendixsec:sem-eqv} for details). Although relying on an LLM judge, correctness is determined via the Lean proof, not the LLM's judgment alone. TC and SC are reported as \textit{pass@k}, i.e., over the top-$k$ generated candidates, for $k = 1,2,4,8,16,32$.


\subsection{State-of-the-Art Baselines}

\textbf{SoTA for NL/FL Cross-Modal Retrieval.} To our knowledge, no existing model jointly embeds theorems and proofs in NL and FL. Pre-trained encoders alone do not yield embeddings suitable for meaningful cross-modal retrieval. To illustrate, we evaluate two \textit{\textbf{SoTA Encoders}}: Qwen3-Embedding-8B~\citep{zhang2025qwen3} and E5-Mistral-7B-Instruct~\citep{wang2024improving}. Qwen3 allows user-defined output dimensions up to 4096, and we use 512 to match our joint embedding model, while E5-Mistral (used by LeanSearch-PS~\citep{shen2025real}) has a fixed dimension of 4096. We also include a \textit{\textbf{Baseline Encoder}}, all-MiniLM-L6-v2~\citep{sentence_transformers}, used for the NL encoder in \ourTool{}, producing 384-dim embeddings. All these encoders treat theorem+proof pairs as plain text, ignoring the DAG structure of FL proofs, which \ourTool{} explicitly leverages. Retrieval is performed via cosine similarity in the respective embedding spaces.

\textbf{SoTA Tools for Proof Auto-Formalization.} Existing proof auto-formalization tools include DSP~\citep{jiang2022draft}, which supports only Isabelle, making comparison infeasible since we target Lean~4. Another recent tool, FormL4~\citep{lu2024process}, has not yet released its trained model. We therefore compare three categories of tools. First, we consider four \textit{\textbf{foundation models}}, including GPT-5-mini~\citep{openai2025gpt5} and the Gemini-2.5 variants, Flash-Lite, Flash, and Pro~\citep{comanici2025gemini}. Next, we evaluate seven \textit{\textbf{AFPS LLMs}} that generate FL proofs with tactics: DeepSeek-Prover-V1.5-RL~\citep{xin2024deepseek15}, STP\_model\_Lean\_0320~\citep{dong2025beyond}, Goedel-Prover-SFT~\citep{lin2025goedel}, Leanabell-Prover-V2-KM~\citep{zhang2025leanabell}, and three Kimina-Prover variants~\citep{wang2025kimina} (RL-1.7B, Distill-8B, 72B). Finally, we evaluate two \textit{\textbf{auto-formalization LLMs}}: Kimina-Autoformalizer-7B~\citep{wang2025kimina} and Herald-Translator~\citep{gao2024herald}.


\subsection{Training and Evaluation Setup}
\label{subsec:expt-setup}

We train our joint embedding model on 90\% (35,056 instances) of \textsc{NuminaMath-Lean-PF} and evaluate it on the remaining 3,895 instances. The split is domain-stratified across all mathematical areas, ensuring hard negatives in the test set. The train split serves as a database $\mathcal{D}$ of FL theorem+proof pairs. \ourTool{}, built on Kimina-Prover-RL-1.7B, is SFT-tuned for NL-to-FL translation using $\big(M_{\text{NL}},M_{\text{FL}}\big)$ from \textsc{NuminaMath-Lean-PF}, with the joint embedding model retrieving the top-5 relevant FL proofs from $\mathcal{D}$ for retrieval-augmented SFT and inference. Inference-time iterative proof repair is applied, and the model is evaluated on \textsc{miniF2F-Test-PF}. See Appendices~\ref{appendixsec:reproducibility}–\ref{appendixsec:hyperparams} for implementation and training details, and Appendix~\ref{appendixsec:example} for an example inference.

The \textit{SoTA Tools} are evaluated in three settings: (a) \textit{zero-shot}, with no in-context I/O examples; (b) \textit{random few-shot}, with five randomly selected in-context examples; and (c) \textit{text-based retrieval few-shot}, where the top-5 FL theorem+proof pairs are retrieved from $\mathcal{D}$ via Qwen3-Embedding-8B and paired with their NL counterparts as in-context examples. Further, we evaluate a \textit{SoTA Two-Step} setting: a theorem-only auto-formalizer (T1) first translates $T_{\text{NL}}$ to $\widetilde{T}_{\text{FL}}$, and an AFPS LLM (T2) then generates $\langle \widetilde{T}_{\text{FL}}, \widetilde{P}_{\text{FL}}\rangle$ from $\widetilde{T}_{\text{FL}}$, both in zero-shot. For \textit{pass@k}, we sample the top-$k$ from T1, select one that is TC and judged equivalent to $T_{\text{NL}}$ by Gemini-2.5-Pro (SC is not used as the gold $T_{\text{FL}}$ is withheld until the pipeline finishes), and then generate the top-$k$ proof candidates from T2.

\subsection{Experimental Results}
\label{sec:exptresults}

\begin{table*}[t]
\centering
\renewcommand{\arraystretch}{1} 
\setlength{\tabcolsep}{4pt} 
\caption{\textbf{NL/FL Cross-Modal Retrieval Performance.} Retrieval performance of \textit{SoTA} and \textit{Baseline} encoders versus \ourTool{}'s joint embedding. ({\small \texttt{\textbf{Q1}}}: 25\textsuperscript{th} \%tile, {\small \texttt{\textbf{M}}}: median, {\small \texttt{\textbf{Q3}}}: 75\textsuperscript{th} \%tile)}
\vspace{-2mm}
\resizebox{\textwidth}{!}{%
\begin{tabular}{l l ccccc  c  cccccc  ccccc c}
\toprule
\multirow{2}{*}{\textbf{Method}} & \multirow{2}{*}{\shortstack{\textbf{Retrieval} \\ \textbf{Direction}}} & 
\multicolumn{5}{c}{\textbf{Recall Rate @ K (\%) $\bm{\uparrow}$}} & \multirow{2}{*}{\textbf{MRR} $\bm{\uparrow}$} & &
\multicolumn{5}{c}{\textbf{Cos. Similarity of top-$\mathbf{K}$ Retrieved} $\bm{\uparrow}$} & 
\multicolumn{5}{c}{\textbf{Cos. Similarity of NOT Retrieved} $\bm{\downarrow}$} & \multirow{2}{*}{\shortstack{\textbf{Gap} $\bm{\uparrow}$ \\ \textbf{(mMG)}}} \\
\cmidrule(lr){3-7} \cmidrule(lr){10-14} \cmidrule(lr){15-19}
 & & $\mathbf{K}$\textbf{=}$\mathbf{1}$ & $\mathbf{K}$\textbf{=}$\mathbf{5}$ & $\mathbf{K}$\textbf{=}$\mathbf{10}$ & $\mathbf{K}$\textbf{=}$\mathbf{20}$ & $\mathbf{K}$\textbf{=}$\mathbf{50}$ & & & $\mathbf{K}$\textbf{=}$\mathbf{1}$ & $\mathbf{K}$\textbf{=}$\mathbf{5}$ & $\mathbf{K}$\textbf{=}$\mathbf{10}$ & $\mathbf{K}$\textbf{=}$\mathbf{20}$ & $\mathbf{K}$\textbf{=}$\mathbf{50}$ & $\mathbf{K}$\textbf{=}$\mathbf{1}$ & $\mathbf{K}$\textbf{=}$\mathbf{5}$ & $\mathbf{K}$\textbf{=}$\mathbf{10}$ & $\mathbf{K}$\textbf{=}$\mathbf{20}$ & $\mathbf{K}$\textbf{=}$\mathbf{50}$ \\
\midrule
\multirow{3}{*}{\makecell{Qwen3-Embedding-8B \\ (\textit{SoTA Encoder}, \\ 8B params)}} & NL $\to$ FL & \underline{46.75} & \underline{71.96} & \underline{82.49} & \underline{87.04} & \underline{93.34} & \underline{0.567} & \shortstack{{\small \texttt{\textbf{Q1}:}} \\ {\small \texttt{\textbf{M}:}} \\ {\small \texttt{\textbf{Q3}:}}} &
\shortstack{0.74 \\ 0.74 \\ 0.74} & \shortstack{0.67 \\ \underline{0.68} \\ 0.70} & \shortstack{0.62 \\ 0.63 \\ 0.66} & \shortstack{0.60 \\ \underline{0.62} \\ 0.63} & \shortstack{0.57 \\ \underline{0.58} \\ 0.60} &
\shortstack{0.317 \\ 0.362 \\ 0.410} & \shortstack{0.317 \\ 0.362 \\ 0.410} & \shortstack{0.317 \\ 0.362 \\ 0.410} & \shortstack{0.317 \\ 0.362 \\ 0.409} & \shortstack{0.317 \\ 0.362 \\ 0.409} & 0.29 \\
\noalign{\vskip 2pt}
\cdashline{2-20}
\noalign{\vskip 5pt}
 & FL $\to$ NL & \underline{44.68} & \underline{68.26} & \underline{79.79} & \underline{83.78} & \underline{87.36} & \underline{0.506} & \shortstack{{\small \texttt{\textbf{Q1}:}} \\ {\small \texttt{\textbf{M}:}} \\ {\small \texttt{\textbf{Q3}:}}} &
\shortstack{0.76 \\ 0.76 \\ 0.76} & \shortstack{0.66 \\ 0.67 \\ 0.68} & \shortstack{0.63 \\ 0.63 \\ 0.65} & \shortstack{0.62 \\ \underline{0.63} \\ 0.64} & \shortstack{0.59 \\ \underline{0.60} \\ 0.62} &
\shortstack{0.309 \\ 0.362 \\ 0.418} & \shortstack{0.309 \\ 0.362 \\ 0.418} & \shortstack{0.309 \\ 0.362 \\ 0.418} & \shortstack{0.309 \\ 0.362 \\ 0.418} & \shortstack{0.308 \\ 0.362 \\ 0.417} & 0.30 \\
\noalign{\vskip 2pt}
\cdashline{1-20}
\noalign{\vskip 5pt}
\multirow{3}{*}{\makecell{E5-Mistral-7B-Instruct \\ (\textit{SoTA Encoder}, \\ 7B params)}} & NL $\to$ FL & 35.60 & 53.22 & 60.22 & 67.82 & 77.18 & 0.441 & \shortstack{{\small \texttt{\textbf{Q1}:}} \\ {\small \texttt{\textbf{M}:}} \\ {\small \texttt{\textbf{Q3}:}}} &
\shortstack{0.86 \\ \textbf{0.86} \\ 0.86} & \shortstack{0.83 \\ \textbf{0.83} \\ 0.84} & \shortstack{0.82 \\ \textbf{0.83} \\ 0.83} & \shortstack{0.82 \\ \textbf{0.82} \\ 0.83} & \shortstack{0.81 \\ \textbf{0.81} \\ 0.82} &
\shortstack{0.711 \\ 0.730 \\ 0.749} & \shortstack{0.711 \\ 0.730 \\ 0.749} & \shortstack{0.711 \\ 0.730 \\ 0.749} & \shortstack{0.711 \\ 0.730 \\ 0.749} & \shortstack{0.710 \\ 0.730 \\ 0.749} & 0.10\\
\noalign{\vskip 2pt}
\cdashline{2-20}
\noalign{\vskip 5pt}
 & FL $\to$ NL & 30.27 & 41.93 & 46.41 & 50.83 & 57.88 & 0.359 & \shortstack{{\small \texttt{\textbf{Q1}:}} \\ {\small \texttt{\textbf{M}:}} \\ {\small \texttt{\textbf{Q3}:}}} &
\shortstack{0.87 \\ \textbf{0.87} \\ 0.87} & \shortstack{0.85 \\ \textbf{0.85} \\ 0.86} & \shortstack{0.84 \\ \textbf{0.85} \\ 0.85} & \shortstack{0.84 \\ \textbf{0.84} \\ 0.85} & \shortstack{0.83 \\ \textbf{0.83} \\ 0.84} &
\shortstack{0.704 \\ 0.735 \\ 0.760} & \shortstack{0.704 \\ 0.735 \\ 0.760} & \shortstack{0.704 \\ 0.735 \\ 0.760} & \shortstack{0.704 \\ 0.735 \\ 0.760} & \shortstack{0.704 \\ 0.734 \\ 0.759} & 0.11\\
\noalign{\vskip 2pt}
\cdashline{1-20}
\noalign{\vskip 5pt}
\multirow{3}{*}{\makecell{all-MiniLM-L6-v2 \\ (\textit{Baseline Encoder}, \\ 22.7M params)}} & NL $\to$ FL & 16.06 & 30.93 & 38.63 & 47.31 & 60.95 & 0.237 & \shortstack{{\small \texttt{\textbf{Q1}:}} \\ {\small \texttt{\textbf{M}:}} \\ {\small \texttt{\textbf{Q3}:}}} &
\shortstack{0.60 \\ 0.60 \\ 0.60} & \shortstack{0.52 \\ 0.54 \\ 0.55} & \shortstack{0.50 \\ 0.51 \\ 0.53} & \shortstack{0.47 \\ 0.49 \\ 0.51} & \shortstack{0.44 \\ 0.45 \\ 0.58} &
\shortstack{0.147 \\ \underline{0.210} \\ 0.274} & \shortstack{0.147 \\ \underline{0.210} \\ 0.274} & \shortstack{0.147 \\ \underline{0.210} \\ 0.273} & \shortstack{0.147 \\ \underline{0.209} \\ 0.273} & \shortstack{0.146 \\ \underline{0.208} \\ 0.271} & \underline{0.31}\\
\noalign{\vskip 2pt}
\cdashline{2-20}
\noalign{\vskip 5pt}
 & FL $\to$ NL & 26.35 & 45.12 & 54.28 & 63.75 & 75.54 & 0.355 & \shortstack{{\small \texttt{\textbf{Q1}:}} \\ {\small \texttt{\textbf{M}:}} \\ {\small \texttt{\textbf{Q3}:}}} &
\shortstack{0.58 \\ 0.58 \\ 0.58} & \shortstack{0.52 \\ 0.53 \\ 0.54} & \shortstack{0.50 \\ 0.51 \\ 0.53} & \shortstack{0.47 \\ 0.49 \\ 0.51} & \shortstack{0.44 \\ 0.46 \\ 0.48} &
\shortstack{0.142 \\ \underline{0.208} \\ 0.278} & \shortstack{0.142 \\ \underline{0.208} \\ 0.278} & \shortstack{0.142 \\ \underline{0.208} \\ 0.277} & \shortstack{0.142 \\ \underline{0.207} \\ 0.277} & \shortstack{0.141 \\ \underline{0.206} \\ 0.275} & \underline{0.31}\\
\midrule
\multirow{2}{*}{\makecell{\textbf{\ourTool{}} \\ (\textit{Proposed}, \\ 22.7M + 218M + 1M \\params)}} & NL $\to$ FL & \textbf{52.83} & \textbf{79.81} & \textbf{87.06} & \textbf{91.49} & \textbf{95.08} & \textbf{0.650} & \shortstack{{\small \texttt{\textbf{Q1}:}} \\ {\small \texttt{\textbf{M}:}} \\ {\small \texttt{\textbf{Q3}:}}} & 
\shortstack{ 0.76 \\  \underline{0.76} \\  0.76} & 
\shortstack{ 0.66 \\  \underline{0.68} \\  0.71} & 
\shortstack{ 0.62 \\  \underline{0.64} \\  0.68} & 
\shortstack{ 0.57 \\  0.60 \\  0.64} & 
\shortstack{ 0.49 \\  0.52 \\ 0.58} & 
\shortstack{ -0.134 \\  \textbf{-0.009} \\  0.123} & 
\shortstack{ -0.135 \\  \textbf{-0.010} \\  0.123} & 
\shortstack{ -0.135 \\  \textbf{-0.010} \\ 0.122} & 
\shortstack{ -0.135 \\  \textbf{-0.010} \\  0.121} & 
\shortstack{ -0.136 \\  \textbf{-0.012} \\  0.117} & \textbf{0.65}\\
\noalign{\vskip 2pt}
\cdashline{2-20}
\noalign{\vskip 5pt}
 & FL $\to$ NL & \textbf{51.23} & \textbf{78.77} & \textbf{86.18} & \textbf{90.50} & \textbf{94.83} & \textbf{0.635} & \shortstack{{\small \texttt{\textbf{Q1}:}} \\ {\small \texttt{\textbf{M}:}} \\ {\small \texttt{\textbf{Q3}:}}} & 
\shortstack{ 0.77 \\  \underline{0.77} \\  0.77} & \shortstack{ 0.67 \\  \underline{0.69} \\  0.71} & \shortstack{ 0.62 \\  \underline{0.64} \\  0.68} & \shortstack{ 0.57 \\  0.60 \\  0.64} & \shortstack{ 0.49 \\  0.52 \\  0.58} &
\shortstack{ -0.135 \\  \textbf{-0.009} \\  0.124} & \shortstack{-0.135 \\  \textbf{-0.009} \\  0.123} & \shortstack{ -0.135 \\  \textbf{-0.009} \\  0.122} & \shortstack{ -0.135 \\  \textbf{-0.011} \\  0.121} & \shortstack{ -0.136 \\  \textbf{-0.012} \\  0.117} & \textbf{0.65}\\
\bottomrule
\end{tabular}%
}
\vspace{-4mm}
\label{tab:retrieval-results}
\end{table*}

\textbf{NL/FL Cross-Modal Retrieval.} Table~\ref{tab:retrieval-results} compares \ourTool{}'s joint embedding model with the two \textit{SoTA Encoders} and the \textit{Baseline Encoder}. Since \ourTool{} is obtained by contrastively training the NL encoder together with an FL encoder, all improvements are reported relative to the original NL encoder (the Baseline Encoder). \ourTool{} achieves consistently higher recall rates across all top-$K$ values: for NL$\to$FL, it yields $3.28\times$ gain for NL$\to$FL and $1.94\times$ for FL$\to$NL at $K=1$. MRR also improves by $2.74\times$ and $1.79\times$ for NL$\to$FL and FL$\to$NL, respectively. This indicates that \ourTool{} more frequently retrieves the correct cross-modal counterpart among the highest-ranked results. For the NL and FL embeddings by \ourTool{}, the median ({\small \texttt{\textbf{M}}}) cosine similarity of retrieved cross-modal instances averages $0.64$ across top-$K$ ($K=1,5,10,20,50$) in both directions, showing that select cross-modal theorem+proof pairs are tightly clustered. In contrast, non-retrieved instances are much farther apart, averaging $-0.01$. Compared to the Baseline, retrieved-pair similarities increase by $23\%$, while non-retrieved similarities decrease by $104\%$.  

\vspace{-1.5mm}
\ourTool{} outperforms the \textit{SoTA Encoders} in recall and MRR, with a much higher mMG despite being 32$\times$ smaller, showing a clearer separation between retrieved and non-retrieved items. For NL$\to$FL, E5-Mistral-7B-Instruct attains an mMG of $0.10$, while Qwen3-Embedding-8B reaches $0.29$. We believe these low values arise because: \textit{first}, these QA-oriented encoders capture coarse domain-level signals rather than fine-grained mathematical semantics, so most mathematical texts cluster together; \textit{second}, as plain-text, non-DAG-aware encoders, they rely on superficial lexical cues (keywords), but in Lean many proofs share the same tactics, making keyword overlap non-discriminative. In contrast, \ourTool{} leverages the DAG to distinguish proofs, achieving an mMG of $0.65$. Overall, encoding Lean proofs via linearized DAG traversals and contrastive alignment with a DAG-aware FL encoder yield an effective joint embedding space, where equivalent NL-FL pairs cluster tightly and inequivalent ones remain well separated. This enables reliable retrieval of the most relevant FL demonstrations to condition the LLM during auto-formalization. 

\noindent
\tikz[overlay,remember picture]{
    \draw[line width=0.2pt]
        ($(current page.north) + (-4.2cm, -3.81cm)$)
        --
        ($(current page.south) + (-4.2cm, 17.85cm)$);

    \draw[line width=0.2pt]
        ($(current page.north) + (-1.15cm, -3.81cm)$)
        --
        ($(current page.south) + (-1.15cm, 17.85cm)$);

    \draw[line width=0.2pt]
        ($(current page.north) + (-0.42cm, -3.81cm)$)
        --
        ($(current page.south) + (-0.42cm, 17.85cm)$);

    \draw[line width=0.2pt]
        ($(current page.north) + (3.05cm, -4.42cm)$)
        --
        ($(current page.south) + (3.05cm, 17.85cm)$);

    \draw[line width=0.2pt]
        ($(current page.north) + (6.26cm, -4.42cm)$)
        --
        ($(current page.south) + (6.26cm, 17.85cm)$);
}

\vspace{-5mm}

\vspace{-2mm}
\textbf{Proof Auto-Formalization.} Table~\ref{tab:pf-af-results} reports the proof auto-formalization performance of 13 SoTA LLM-based tools. \textit{Theorem auto-formalization LLMs} achieve 0\% TC and SC across all pass@k. These models are designed to formalize theorem statements only, leaving proofs as \lstinline{sorry}. They lack knowledge of proof DAGs and tactics, making them unsuitable for end-to-end proof auto-formalization. Among \textit{foundation models}, Gemini-2.5-Flash-Lite achieves 2.87\% SC and 21.31\% TC at pass@32, which increase to 4.10\% SC, 18.44\% TC for Flash and 8.61\% SC, 31.56\% TC for Pro. GPT-5-mini attains 9.02\% TC and 34.84\% SC at pass@32. They struggle with the strict syntax and semantics of specialized FLs like Lean, which are underrepresented in their training data. The \textit{SoTA Two-Step} achieves 43.44\% SC and 59.43\% TC, but it is prone to cascading errors: an incorrect FL theorem from the first model causes the second to produce a semantically incorrect theorem+proof pair. Among the \textit{SoTA AFPS LLMs}, Kimina-Prover-72B achieves the strongest zero-shot performance at pass@32, with 46.31\% SC and 79.51\% TC. We build \ourTool{} on top of a smaller variant, Kimina-Prover-RL-1.7B, by retrieving five relevant FL proofs via NL/FL cross-modal retrieval and using them for retrieval-augmented SFT and inference, along with iterative proof repair. \ourTool{} surpasses the zero-shot performance of Kimina-Prover-RL-1.7B by +22.54\% SC and +20.49\% TC, and its random few-shot performance by +31.14\% SC and +1.64\% TC.

\begin{wrapfigure}[16]{r}[0pt]{0.38\textwidth} 
\vspace{-13mm}
  \centering
    \includegraphics[width=0.92\linewidth]{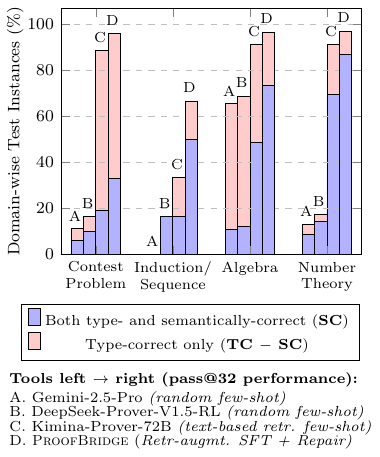}%
    \vspace{-3mm}
\caption{\textbf{Category-wise Results.} Proof auto-formalization performance across mathematical domains.}
  \label{fig:perfByDomain}
\end{wrapfigure}

\vspace{-1.5mm}
In Figure~\ref{fig:perfByDomain}, we present the pass@32 performance across mathematical domains. Following the taxonomy by~\cite{zheng2021minif2f}, the benchmark includes 6 induction/sequence (2.46\%), 69 number-theory (28.28\%), 90 algebra (36.89\%), and 79 contest problems (32.38\%) sourced from AIME, AMC, and IMO. \ourTool{} performs best on number theory, achieving over 85\% SC, while contest problems remain the most challenging, reaching only about 35\% SC.


\begin{table*}[!t]
\centering
\renewcommand{\arraystretch}{1.05}
\setlength{\tabcolsep}{2pt}
\caption{\textbf{Proof Auto-Formalization Performance.} 
Comparison of LLM-based tools on \emph{Semantic Correctness (SC)} and \emph{Type Correctness (TC)} across pass@$k$ metrics ($k \in \{1,2,4,8,16,32\}$).}
\vspace{-2mm}
\resizebox{\textwidth}{!}{%
\begin{tabular}{c l ccccccc c cccccc}
\toprule
\multirow{2}{*}{\textbf{Setting}} & \multirow{2}{*}{\textbf{LLM/Tool}} & 
\multicolumn{6}{c}{\textbf{Semantic Correctness (SC) (\%) $\bm{\uparrow}$}} & 
& \multicolumn{6}{c}{\textbf{Type Correctness (TC) (\%) $\bm{\uparrow}$}} \\
\cmidrule(lr){3-8} \cmidrule(lr){10-15}
 & & \textbf{pass@1} & \textbf{pass@2} & \textbf{pass@4} & \textbf{pass@8} & \textbf{pass@16} & \textbf{pass@32} & 
 & \textbf{pass@1} & \textbf{pass@2} & \textbf{pass@4} & \textbf{pass@8} & \textbf{pass@16} & \textbf{pass@32} \\
\midrule
\multirow{3}{*}{\makecell{\textit{SoTA Tools} \\ (zero-shot)}} & Kimina-Prover-RL-1.7B & 9.02 & 13.93 & 22.54 & 30.33 & 35.25 & 40.16 & & 26.23 & 41.80 & 56.15 & 62.70 & 68.03 & 75.00 \\
& Kimina-Prover-Distill-8B & 10.66 & 18.85 & 23.77 & 32.38 & 37.70 & 41.80 &  & 27.05 & 43.03 & 58.20 & 63.52 & 72.95 & 75.82 \\
& Kimina-Prover-72B & 12.70 & \underline{21.31} & 25.00 & 34.84 & 38.52 & 43.03 &  & 30.33 & 45.08 & 61.07 & 69.26 & 75.41 & 79.51 \\
\midrule
\multirow{11}{*}{\makecell{\textit{SoTA Tools} \\ (random \\few-shot)}} & Kimina-Autoformalizer-7B & 0.00 & 0.00 & 0.00 & 0.00 & 0.00 & 0.00 &  & 0.00 & 0.00 & 0.00 & 0.00 & 0.00 & 0.00 \\
& Herald-Translator & 0.00 & 0.00 & 0.00 & 0.00 & 0.00 & 0.00 &  & 0.00 & 0.00 & 0.00 & 0.00 & 0.00 & 0.00 \\
\noalign{\vskip 1pt}
\cdashline{2-15}
\noalign{\vskip 1pt}
& Gemini-2.5-Flash-Lite & 0.00 & 0.00 & 0.00 & 0.00 & 1.23 & 2.87 &  & 0.82 & 4.51 & 9.02 & 13.93 & 18.03 & 21.31 \\
& Gemini-2.5-Flash & 0.00 & 0.82 & 2.05 & 2.86 & 3.28 & 4.10 &  & 2.45 & 4.92 & 7.38 & 12.30 & 15.98 & 18.44 \\
& Gemini-2.5-Pro & 1.23 & 1.23 & 3.28 & 4.92 & 6.97 & 8.61 & & 9.84 & 13.52 & 18.85 & 23.77 & 29.10 & 31.56 \\
& GPT-5-mini & 0.41 & 1.23 & 4.51 & 6.97 & 7.38 & 9.02 & & 4.92 & 9.43 & 20.08 & 28.28 & 32.38 & 34.84 \\
\noalign{\vskip 1pt}
\cdashline{2-15}
\noalign{\vskip 1pt}
& DeepSeek-Prover-V1.5-RL & 3.69 & 6.15 & 8.61 & 9.02 & 11.07 & 12.30 & & 8.20 & 14.34 & 19.67 & 24.18 & 28.68 & 35.66 \\
& STP\_model\_Lean\_0320 & 4.51 & 6.56 & 8.20 & 9.84 & 11.48 & 13.11 & & 12.70 & 18.03 & 23.36 & 28.69 & 33.61 & 39.34 \\
& Goedel-Prover-SFT & 4.92 & 5.33 & 7.38 & 8.20 & 12.70 & 16.39 & & 13.52 & 17.21 & 25.41 & 31.56 & 36.88 & 42.21 \\
& Leanabell-Prover-V2-KM & 6.97 & 9.43 & 10.66 & 13.52 & 15.16 & 18.03 & & 16.80 & 21.31 & 27.87 & 37.30 & 41.39 & 50.41 \\
& Kimina-Prover-RL-1.7B & 6.15 & 12.30 & 17.62 & 22.13 & 27.46 & 31.56 & & 26.23 & 42.21 & 60.66 & 74.18 & \underline{88.11} & \underline{93.85} \\
& Kimina-Prover-Distill-8B & 7.38 & 11.89 & 16.39 & 24.18 & 28.69 & 32.38 &  & 24.59 & 44.26 & 61.89 & 75.00 & 85.25 & 89.34 \\
& Kimina-Prover-72B & 10.25 & 13.93 & 18.85 & 25.00 & 31.35 & 37.30 &  & 30.74 & 45.49 & 62.70 & \underline{77.87} & 86.89 & 91.39 \\
\midrule
\multirow{3}{*}{\makecell{\textit{SoTA Tools} \\ (text-based retrieval\\ few-shot)}} & Kimina-Prover-RL-1.7B & 6.15 & 12.70 & 18.85 & 22.54 & 28.28 & 32.78 &  & 26.63 & 43.39 & 58.68 & 70.66 & 86.36 & 89.75 \\
& Kimina-Prover-Distill-8B & 8.61 & 13.52 & 21.72 & 28.28 & 34.02 & 36.07 &  & 28.28 & 46.28 & 59.92 & 71.90 & 83.47 & 86.07 \\
& Kimina-Prover-72B & 12.29 & 14.34 & 24.59 & 29.51 & 37.70 & 44.26 &  & \underline{31.15} & 45.08 & \textbf{64.88} & 75.62 & 86.78 & 88.93 \\
\midrule
\multirow{1}{*}{\makecell{\textit{SoTA Two-Step}}} & \multirow{1}{*}{\makecell{Herald-Translator $\to$ Kimina-Prover-Distill-8B}} & \underline{14.75} & 19.26 & 27.05 & 33.20 & 38.52 & 43.44 &  & 30.33 & 32.79 & 43.03 & 48.36 & 54.10 & 59.43 \\
\midrule
\multirow{3}{*}{\makecell{\textit{Our Tool}}} & \textbf{\ourTool{}} (SFT only) & 6.97 & 13.52 & 19.26 & 24.18 & 29.92 & 34.84 & & 27.87 & 45.90 & 60.66 & 66.39 & 72.13 & 78.69 \\
& \textbf{\ourTool{}} (Retrieval-augmt. SFT) & 13.11 & 20.90 & \underline{27.87} & \underline{35.66} & \underline{47.95} & \underline{55.33} & & 29.92 & \underline{46.31} & 60.25 & 71.31 & 83.20 & 89.75 \\
& \textbf{\ourTool{}} (Retrieval-augmt. SFT + Repair) & \textbf{16.39} & \textbf{25.41} & \textbf{29.51} & \textbf{37.70} & \textbf{50.41} & \textbf{62.70} & & \textbf{32.79} & \textbf{47.13} & \underline{64.75} & \textbf{78.69} & \textbf{90.16} & \textbf{95.49} \\
\bottomrule
\end{tabular}%
}
\vspace{-4mm}
\label{tab:pf-af-results}
\end{table*}




\subsection{Ablation Studies}

\setlength{\columnsep}{7pt}
\setlength{\fboxsep}{0pt} 

To assess the effect of in-context examples, Table~\ref{tab:pf-af-results} reports zero-shot, random few-shot, and text-based retrieval few-shot performance for three Kimina-Prover variants (72B, Distill-8B, and RL-1.7B). From the pass@32 results, Kimina-Prover-RL-1.7B achieves $40.16$\% SC and $75.00$\% TC in the zero-shot setting. When random examples are added, SC drops to $31.56$\% while TC rises to $93.85$\%, with similar trends across variants. This indicates that random examples improve TC (syntax) but hurt SC by causing the model to hallucinate semantically misaligned proofs. With text-based retrieval via Qwen3-Embedding-8B, SC rises to $32.38$\% but TC declines, likely because QA-based retrieval favors proofs with similar tactics, reducing tactic diversity. This highlights the need for retrieving semantically relevant examples via a DAG-aware encoder, as in \ourTool{}.

To quantify the contribution of each component of \ourTool{}, we perform an ablation over three variants. \textbf{\ourTool{} (SFT)}, fine-tuned on labeled NL-FL pairs and evaluated in the few-shot setting with semantically relevant examples via our joint-embedding model, improves SC by $+2.06$\% but reduces TC by $-11.06$\% relative to Kimina-Prover-RL-1.7B (text-based retrieval few-shot). Next, in \textbf{\ourTool{} (Retrieval-augmented SFT)}, we fine-tune the LLM with semantically relevant FL proofs included in the input, achieving $+22.55$\% SC. \textbf{\ourTool{} (Retrieval-augmented SFT + Repair)}, adding iterative proof repair, yields the best results: $+29.92$\% SC and $+5.74$\% TC over the same baseline. Relative to Kimina-Prover-RL-1.7B (random few-shot), the improvements are $+31.14$\% SC and $+1.64$\% TC.

\section{Conclusion}
\label{sec:concl}

We present \ourTool{}, a unified framework for NL-to-Lean proof auto-formalization that translates both theorems and proofs end-to-end. At its core is a joint embedding model of NL and FL that encodes Lean proof DAGs, capturing tactic sequences and dependency structures. It enables highly effective cross-modal retrieval of semantically relevant FL proofs. These retrieved proofs act as demonstrations, guiding retrieval-augmented fine-tuning of an LLM. An iterative verifier-guided repair loop further refines generated proofs by combining Lean type-checking with semantic equivalence checking to ensure correctness. Evaluated on \textsc{miniF2F-Test-PF}, \ourTool{} significantly outperforms state-of-the-art LLMs in both semantic correctness (by bi-directional equivalence proving) and type correctness, demonstrating that integrating structured embeddings, retrieval guidance, and verifier feedback leads to more reliable proof auto-formalization.

\clearpage
\bibliography{iclr2026_conference}
\bibliographystyle{iclr2026_conference}

\clearpage
\appendix
\section{Appendix}

\subsection{The Use of Large Language Models (LLMs)}

LLMs did \textit{not} play a significant role in either the research ideation or the writing of this paper. Their use was limited to correcting minor grammatical issues and typographical errors.

\subsection{Reproducibility and Datasets}
\label{appendixsec:reproducibility}

All implementations in this work, including dataset construction, model training, cross-modal retrieval, inference, and evaluation, use Python 3.12.10 and Lean v4.15.0. Experiments were conducted on a high-performance AlmaLinux 9.5 (Teal Serval) cluster with a single Intel Xeon Platinum 8480+ CPU (32 cores, 2.0-4.0 GHz), 251 GiB of RAM, and one NVIDIA H100 GPU. Our full codebase, including scripts for dataset generation, model fine-tuning, and inference across both GPU- and API-based setups, is available at \url{https://github.com/PrithwishJana/ProofBridge}. The repository is well documented and includes instructions for creating a Python virtual environment, along with a comprehensive README detailing library dependencies, dataset formats, and step-by-step instructions to replicate the entire data pipeline and experimental workflow.

Through our repository, we release \textsc{NuminaMath-Lean-PF} and \textsc{miniF2F-Test-PF}, following the respective licenses (Apache License 2.0 or MIT License) of their source datasets. Note that our \textsc{NuminaMath-Lean-PF} dataset is carefully refined and extended from NuminaMath-LEAN~\citep{wang2025kimina}, while \textsc{miniF2F-Test-PF} is curated by combining two versions of miniF2F-test~\citep{zheng2021minif2f} from \cite{AIMO_minif2f_test_2025} and \cite{yangky11_miniF2F_lean4_2025}. The original datasets were built by collecting problems from web sources, forums, and online documents, including worldwide competitions such as AMC, AIME, and IMO, as well as public K–12 exam papers. For each problem, the corresponding Lean theorem and proof were produced either by human annotators or via an LLM-based auto-formalizer. Formalizing large-scale proofs often requires an agile methodology~\citep{jana2020essence,jana2025real}, dividing proofs into small, manageable lemmas that can be handled by different human annotators. Many large-scale Lean projects now use tools such as \texttt{leanblueprint}~\citep{leanblueprint}, which serve as planning and coordination resources, enabling contributors to track progress, visualize dependencies, and identify remaining formalization tasks. Because the pipeline aggregates material from problem authors, human formalizers, LLM-based auto-formalizers, textbooks, competition archives, and exams, it is important to ensure proper privacy, licensing, and copyright compliance~\citep{jana2013efficient}. 

\subsection{Training and Inference Hyperparameters}
\label{appendixsec:hyperparams}

\textbf{NL/FL Cross-Modal Retrieval.} We train two dense encoders to embed NL and FL theorem+proof pairs into a shared semantic space. The NL encoder is initialized from \texttt{all-MiniLM-L6-v2}~\citep{sentence_transformers}, while the FL encoder builds on LeanDojo's~\citep{yang2023leandojo} \texttt{ByT5}-based proof-state encoder, extended to process a linearized traversal of the proof DAG. Each encoder is equipped with a projection head that maps representations into a shared embedding space of dimension $d=512$, with a dropout rate of $0.1$. During fine-tuning, we update only the top layers of each encoder to retain their pretrained linguistic and structural priors. Specifically, we train the last 3 layers of the NL encoder and the last 2 layers of the FL encoder. We set the maximum token length to 512 for both NL and FL sequences. The model is optimized using our symmetric contrastive objective (Equation~\ref{equation:contrastive-objective}) with temperature $\tau=0.07$, trained using AdamW with a learning rate of $1\times10^{-5}$, weight decay of $0.01$, and a batch size of $32$. Training is run for $10$ epochs with gradient accumulation steps set to $4$. We also enable gradient checkpointing to reduce memory usage during fine-tuning.

\textbf{Proof Auto-Formalization.} \ourTool{} builds on Kimina-Prover-RL-1.7B~\citep{wang2025kimina}, which we further fine-tune for NL→FL translation using paired data $\big(M_{\text{NL}}, M_{\text{FL}}\big)$ from \textsc{NuminaMath-Lean-PF}. During both SFT and inference, our NL/FL cross-modal retrieval model gets the top-5 most relevant FL proofs from $\mathcal{D}$, which are provided as in-context demonstrations to guide Lean proof synthesis. We use the HuggingFace Trainer for supervised fine-tuning with the following settings: a per-device batch size of $8$ and BF16 training enabled. Training is run for $5$ epochs with a learning rate of $5\times 10^{-6}$, cosine decay scheduling, and a warmup ratio of $0.05$. For all SoTA baselines in Table~\ref{tab:pf-af-results}, we compute pass@k using stochastic decoding. Specifically, we run LLM inference with a temperature of $0.6$ and top-p sampling of $0.95$, ensuring sufficient diversity across generated candidates.

\subsection{Semantic Equivalence of Lean theorems}
\label{appendixsec:sem-eqv}

Lean~\citep{moura2021Lean} is a functional programming language in which functions are first-class values, treated like any other data type~\citep{jana2014cpp}. The task of auto-formalization is to convert a mathematical theorem and proof from natural language (NL) into a formal language (FL), such as Lean. When evaluating the performance of such systems, we propose two criteria for evaluation: \emph{type correctness} and \emph{semantic correctness}. Type correctness, which requires that the generated Lean proof is accepted by the Lean type-checker, is straightforward to verify and serves as the standard evaluation metric in the field. In contrast, semantic correctness, which ensures the FL theorem faithfully represents the meaning of the original NL theorem, presents a far greater challenge. To the best of our knowledge, such semantic equivalence has not been systematically evaluated in prior work. This section introduces a novel methodology to address this gap.

While directly measuring the semantic alignment between a NL theorem and a Lean theorem is an unsolved challenge, showing the logical equivalence of two Lean theorems is a tractable task. Our training dataset, \textsc{NuminaMath-Lean-PF}, contains pairs of $\langle T_{\text{NL}}, T_{\text{FL}}\rangle\ $ where most of the $T_{\text{FL}}$ were manually created by experts at Numina. We treat these high-quality $T_{\text{FL}}$ theorems as gold-standard references, assuming they are faithful translations of their $T_{\text{NL}}$ counterparts. This allows us to reduce the intractable problem of verifying a model's generated theorem $\widetilde{T}_{\text{FL}}$ against the original $T_{\text{NL}}$ to the more tractable task of checking for logical equivalence between $\widetilde{T}_{\text{FL}}$ and the gold-standard reference $T_{\text{FL}}$, which can be checked in Lean itself.

To be more specific, we enforce this semantic equivalence check by proving the logical biconditional $\widetilde{T}_{\text{FL}} \leftrightarrow T_{\text{FL}}$ in Lean. Theorems like $\widetilde{T}_{\text{FL}}$ and $T_{\text{FL}}$ are of type \lstinline{Prop} in Lean. The following theorem from Mathlib states that for any two propositions, a logical biconditional between two propositions is itself logically equivalent to their propositional equality: 
\begin{lstlisting}[language=lean]
theorem propext_iff{a b : Prop} :
a = b ↔ (a ↔ b)
\end{lstlisting}

The task thus converts to proving the equality $\widetilde{T}_{\text{FL}} = T_{\text{FL}}$ within Lean. This requires clarifying the specific notion of equality being used, as Lean distinguishes between three primary types: \emph{syntactic}, \emph{definitional}, and \emph{propositional}~\citep{buzzard2022}. 
Syntactic equality is the strictest form of equality in Lean, as it only admits expressions that are structurally identical according to their Abstract Syntax Trees, without any computation or reduction. Definitional equality is a more relaxed form of equality than syntactic equality, where two expressions are considered equal if they compute or reduce to the same normal form. Propositional equality is the weakest form of equality, and also the standard notion of equality used in mathematical theorems. Two terms \lstinline{a, b} are propositionally equal in Lean if we can construct a proof term for the proposition \lstinline{a = b}. 

For our evaluation, we seek to measure how closely a $\widetilde{T}_{\text{FL}}$ matches the $T_{\text{FL}}$. The strictest criterion, syntactic equality, is too restrictive given the current state-of-the-art, as it would fail valid theorems with trivial notational differences. Conversely, full propositional equality can be too permissive; a proof of equivalence can be arbitrarily complex, making it difficult to automate and decide.

Therefore, we adopt a pragmatic compromise: we check for \textbf{definitional equality} supplemented by a form of \textbf{bounded propositional equality}. This means we primarily check if $\widetilde{T}_{\text{FL}}$ and $T_{\text{FL}}$ reduce to the same normal form, but we also permit some propositional equality, provided they can be proven using a collection of tactics  so that their proof complexity is bounded. 

We then leverage Gemini 2.5 Pro as an automated equivalence checker. The model is prompted to synthesize a proof for the biconditional theorem ($\widetilde{T}_{\text{FL}} \leftrightarrow T_{\text{FL}}$), with instructions limiting it to a specific subset of available tactics. This restricted set includes three powerful automated tactics, \lstinline{rfl}, \lstinline{simp}, and \lstinline{ring}, each is designed to discharge a specific class of goals: \lstinline{rfl} for definitional equality, \lstinline{simp} for simplification, and \lstinline{ring} for polynomial identities. 

As definitional equality is our primary target, the equivalence checker first attempts to solve the goal with the \lstinline{rfl} tactic. This single tactic should suffice for the majority of cases. If \lstinline{rfl} fails, the checker then tries \lstinline{simp}. This tactic performs additional simplifications by rewriting the goal using theorems from Mathlib that are tagged for its use. Critically, we use \lstinline{simp} without any arguments. Providing explicit arguments would require a demanding search for the correct lemmas and could introduce unbounded complexity, violating our goal of a bounded proof search. Furthermore, the need for \lstinline{simp} with arguments could imply that the required rewrite is non-trivial, since the default simplification set contains most of the trivial facts~\footnote{It is important to note that the default simp set intentionally excludes lemmas like associativity and commutativity, as they can cause the simplifier to loop indefinitely. However, since these lemmas primarily concern algebraic expressions, they can be handled by the \lstinline{ring} tactic.}. Since our goal is to ensure a close correspondence between $\widetilde{T}_{\text{FL}}$ and $T_{\text{FL}}$, a proof requiring such a targeted rewrite indicates a semantic distance that we classify as a mismatch. The \lstinline{ring} tactic is a valuable complement to the previous tactics as it specializes in proving polynomial equalities. The \lstinline{ring} tactic operates by reducing arithmetic expressions to a canonical normal form. This allows it to prove the equivalence of expressions that are algebraically identical but not definitionally so, such as \lstinline{x^2} and \lstinline{x * x}, which \lstinline{rfl} and default \lstinline{simp} would otherwise fail to solve. 

The three tactics discussed above cover most of the direct equivalences we aim to check. The remaining tactics in our instruction set are designed for a more nuanced case: proving the biconditional when two theorems differ only in their use of auxiliary variables. We observed that human experts and language models may make different but equally valid decisions on whether to introduce an auxiliary variable. We therefore classify such theorems as equivalent. For example, consider the following: 
\begin{lstlisting}[language=lean]
def Prop1 := (∀ (b h v : ℝ), (0 < b ∧ 0 < h ∧ 0 < v) → (v = 1 / 3 * (b * h)) → (b = 30) → (h = 13 / 2) → v = 65)
def Prop2 := (∀ {B h : ℝ}, (B = 30) → (h = 6.5) → (1 / 3) * B * h = 65)
example : Prop1 ↔ Prop2 := by 
  constructor
  · intro
    simp
    ring
  · simp
    intros
    nlinarith
\end{lstlisting}

The main difference between the two propositions \lstinline{Prop1} and \lstinline{Prop2} is the presence of the auxiliary variable \lstinline{v} in one. To prove that such theorems are equivalent, one must typically prove the biconditional by separately proving the implications of both direction. This requires a step-by-step proof construction, and the tactics above are included in our instruction set.

Finally, we note that the LLM judge's role is only to synthesize a biconditional proof under the bounded tactic set described above; the produced proof is then fully type-checked by the Lean kernel, so the SC decision ultimately depends on Lean's verifier (making the metric conservative rather than prone to false positives).

\textbf{Comparison with Existing Semantic Correctness Metrics.} Prior work has proposed similar semantic correctness metrics, including BEq~\citep{liu2025rethinking,wu2025stepfun} and its extensions BEq+~\citep{poiroux2025reliable}. While the general idea behind our SC metric and BEq is similar, both aiming to establish bidirectional equivalence in Lean, the sets of allowed tactics differ. Because the notion of equivalence depends on the permitted tactics, these differences lead to meaningful distinctions between SC and BEq. BEq+ is a reference-based metric inspired by BEq and uses a set of tactics comparable to SC. However, BEq+ is deterministic and CPU-efficient, while SC relies on an LLM-based proof synthesizer. This creates a trade-off: the LLM can capture equivalences beyond the reach of the deterministic procedure, whereas BEq+ provides a reproducible evaluation. 

\vspace{-1mm}
Consider \lstinline{Prop1} and \lstinline{Prop2} above as an example. \lstinline{Prop1} (gold-standard FL theorem from miniF2F-Test-PF) explicitly introduces an auxiliary variable \texttt{v} to denote volume, whereas \lstinline{Prop2} (produced by Kimina-Prover-RL-1.7B) omits the auxiliary variable and substitutes the corresponding formula directly. The tactic set allowed by BEq is not expressive enough to establish equivalence in such cases, so these theorems would not be recognized as equivalent under BEq. Our SC metric, by contrast, was specifically designed to handle such variations, reflecting the fact that human experts may also differ in whether they introduce auxiliary variables. By explicitly handling these variations and using LLM-generated bidirectional proofs that are type-checked, SC provides an evaluation that is both more lenient and faithful in assessing the performance of auto-formalization models.

\subsection{Illustrative Example}
\label{appendixsec:example}

We present an example of an NL theorem+proof pair from \textsc{miniF2F-Test-PF} and compare the \texttt{pass@1} output of proof auto-formalization generated by Kimina-Prover-RL-1.7B in the text-based retrieval few-shot setting with that produced by \ourTool{} (using Retrieval-augmented SFT + Repair). We first show the retrievals of semantically relevant FL theorem+proof pairs from $\mathcal{D}$, followed by the \texttt{pass@1} output proof auto-formalization generated by \ourTool{}. In this example, the output by Kimina-Prover-RL-1.7B is type-correct (TC) but not semantically correct (SC), i.e., it is not bi-directionally equivalent to the gold-standard Lean proof. In contrast, the output by \ourTool{} is both TC and SC.

\tcbset{
  colback=gray!8,
  colframe=gray!90,
  colbacktitle=gray!100,  
  arc=3pt,
  boxrule=0.4pt,
  left=6pt,
  right=6pt,
  top=6pt,
  bottom=6pt,
}

\begin{tcolorbox}[title={\textbf{Input NL theorem+proof pair} (\textsc{miniF2F-Test-PF})}, fontupper=\footnotesize]
\texttt{\textbf{<informal\_theorem>}}

A point $(x,y)$ on the coordinate plane with both coordinates negative is a distance of 6 units from the $x$-axis. It is a distance of 15 units from the point $(8,3)$. It is a distance $\sqrt{n}$ from the origin. What is $n$? Show that it is 52.

\texttt{\textbf{</informal\_theorem>}}\\[6pt]
\texttt{\textbf{<informal\_proof>}}

We know that $y=-6$ from the given information. By the distance formula, we have the equation $\sqrt{(x-8)^2+(-6-3)^2}=15$. Solving, we have \begin{align*}
\sqrt{(x-8)^2+(-6-3)^2}&=15 \\
x^2-16x+64+81&=225 \\
x^2-16x-80&=0 \\
(x-20)(x+4)&=0
\end{align*}Thus, $x+4=0$ or $x-20=0$, so $x=-4$ or $x=20$. $x=-4$ by the given conditions. Thus, our point is $(-4,-6)$ and is a distance of $\sqrt{(-4)^2+(-6)^2}=\sqrt{52}$ units from the origin. $n=52$.

\texttt{\textbf{</informal\_proof>}}
\end{tcolorbox}

\begin{tcolorbox}[title={\textbf{Gold-standard Lean theorem} (\textsc{miniF2F-Test-PF})}, listing only, fontupper=\small]
\vspace{-3mm}
\begin{lstlisting}[language=Lean, basicstyle=\scriptsize\ttfamily]
import Mathlib
import Aesop
set_option maxHeartbeats 0
open BigOperators Real Nat Topology Rat

theorem mathd_algebra_288 (x y : ℝ) (n : NNReal) (h₀ : x < 0 ∧ y < 0) (h₁ : abs y = 6)
    (h₂ : Real.sqrt ((x - 8) ^ 2 + (y - 3) ^ 2) = 15)
    (h₃ : Real.sqrt (x ^ 2 + y ^ 2) = Real.sqrt n) : n = 52 := by
    sorry
\end{lstlisting}
\vspace{-3mm}
\end{tcolorbox}

\begin{tcolorbox}[title={
        \textbf{Output Lean theorem+proof pair by Kimina-Prover-RL-1.7B}\\
        \textnormal{(text-based retrieval few-shot setting)}
    }, listing only, breakable]
\vspace{-3mm}
\begin{lstlisting}[language=Lean, mathescape=true, basicstyle=\scriptsize\ttfamily]
import Mathlib

/- A point (x,y) on the coordinate plane with both coordinates negative is a distance of 6 units from the x-axis. It is a distance of 15 units from the point (8,3). It is a distance \sqrt{n} from the origin. What is n? Show that it is 52.-/

theorem algebra_60927 (n : ℕ) (hn : n ≥ 1) (h : ∃ x y : ℝ, x < 0 ∧ y < 0 ∧ |y| = 6 ∧ $\sqrt{}$((x - 8)^2 + (y - 3)^2) = 15 ∧ $\sqrt{}$(x^2 + y^2) = $\sqrt{}$n) : n = 52 := by
  rcases h with ⟨x, y, hx, hy, h1, h2, h3⟩
  have hy1 : y = -6 := by
    cases' abs_cases y with hy2 hy2
    · linarith
    · linarith
  have h2' : (x - 8)^2 + (y - 3)^2 = 225 := by
    calc
      (x - 8)^2 + (y - 3)^2 = ($\sqrt{}$((x - 8)^2 + (y - 3)^2)) ^ 2 := by rw [Real.sq_sqrt]; positivity
      _ = 15 ^ 2 := by rw [h2]
      _ = (225 : ℝ) := by norm_num
  rw [hy1] at h2'
  have h2x : (x - 8)^2 = 144 := by nlinarith
  have hx1 : x = -4 := by
    nlinarith [h2x, hx]
  have h3' : x^2 + y^2 = (n : ℝ) := by
    calc
      x^2 + y^2 = ($\sqrt{}$(x^2 + y^2)) ^ 2 := by rw [Real.sq_sqrt]; positivity
      _ = ($\sqrt{}$n) ^ 2 := by rw [h3]
      _ = (n : ℝ) := by rw [Real.sq_sqrt]; positivity
  rw [hx1, hy1] at h3'
  have hn1 : (n : ℝ) = 52 := by nlinarith
  have hn2 : n = 52 := by
    exact_mod_cast hn1
  exact_mod_cast hn2
\end{lstlisting}
\vspace{-3mm}
\end{tcolorbox}

Note that, although the Lean theorem generated by Kimina-Prover-RL-1.7B is type-correct (TC), it differs semantically from the gold-standard theorem. The main difference lies in the quantification of variables. In the gold-standard theorem, the variables \(x, y, n\) are universally quantified as explicit arguments, and all hypotheses are stated as direct assumptions; this asserts that for any triple \((x, y, n)\) satisfying the geometric constraints, \(n = 52\). In contrast, the Kimina-Prover-RL-1.7B output universally quantifies \(n\) but existentially quantifies \(x\) and \(y\) within the hypotheses. This more accurately reflects the intended geometric meaning: for a given \(n\) satisfying the distance constraints, there exists a point \((x, y)\) realizing those constraints, and consequently \(n = 52\). Therefore, while both theorems are syntactically valid in Lean, they encode slightly different logical statements. This difference prevents an LLM judge from producing a type-checkable proof of bi-directional equivalence between the two theorems. As a result, the Kimina-Prover-RL-1.7B's output is not semantically correct (SC).

\begin{tcolorbox}[title={
        \textbf{Comparison between the gold-standard theorem and Kimina-Prover-RL-1.7B's output}
    }, listing only, breakable]
\vspace{-3mm}
\begin{lstlisting}[language=Lean, mathescape=true, basicstyle=\scriptsize\ttfamily]
/- Gold-standard theorem -/
theorem mathd_algebra_288 (x y : ℝ) (n : NNReal) (h₀ : x < 0 ∧ y < 0) (h₁ : abs y = 6)
    (h₂ : Real.sqrt ((x - 8) ^ 2 + (y - 3) ^ 2) = 15)
    (h₃ : Real.sqrt (x ^ 2 + y ^ 2) = Real.sqrt n) : n = 52 := by
    sorry

/- Kimina-Prover-RL-1.7B output theorem -/
theorem algebra_60927 (n : ℕ) (hn : n ≥ 1) (h : ∃ x y : ℝ, x < 0 ∧ y < 0 ∧ |y| = 6 ∧ $\sqrt{}$((x - 8)^2 + (y - 3)^2) = 15 ∧ $\sqrt{}$(x^2 + y^2) = $\sqrt{}$n) : n = 52 := by
  sorry
\end{lstlisting}
\vspace{-3mm}
\end{tcolorbox}

Below, we present the relevant FL theorem+proof pairs (demonstrations) from $\mathcal{D}$ retrieved by \ourTool{}, along with their relevance scores.

\begin{tcolorbox}[
    title={\textbf{Lean theorem+proof pairs retrieved by \ourTool{}'s NL/FL Cross-Modal Retrieval}},
    listing only,
    breakable,
    enhanced,
    fontupper=\small
]
\textbf{\#\#\# Relevant Lean theorem+proof 1, with relevance score 0.786764 out of 1.0}

\begin{lstlisting}[language=Lean, basicstyle=\scriptsize\ttfamily]
import Mathlib

/- Find the distance between the points $(2,2)$ and $(-1,-1)$. -/
theorem algebra_13734 (p1 p2 : ℝ × ℝ) (hp1 : p1 = (2, 2)) (hp2 : p2 = (-1, -1)) :
    Real.sqrt ((p1.1 - p2.1)^2 + (p1.2 - p2.2)^2) = 3 * Real.sqrt 2 := by
  rw [hp1, hp2]
  norm_num
  ring
  rw [Real.sqrt_eq_iff_sq_eq] <;> norm_num
  ring
  norm_num
\end{lstlisting}

\textbf{\#\#\# Relevant Lean theorem+proof 2, with relevance score 0.768226 out of 1.0}

\begin{lstlisting}[language=Lean, basicstyle=\scriptsize\ttfamily]
import Mathlib
open Real

/- Prove that the angle (in degrees) between the vectors $(2,5)$ and $(-3,7)$ is $45$. -/
theorem calculus_17161 :
    arccos ((2 * (-3) + 5 * 7) / (sqrt (2 ^ 2 + 5 ^ 2) * sqrt ((-3) ^ 2 + 7 ^ 2))) * 180 / π = 45 := by

  have h1 : (2 * (-3) + 5 * 7 : ℝ) / (sqrt (2 ^ 2 + 5 ^ 2) * sqrt ((-3) ^ 2 + 7 ^ 2)) = Real.sqrt 2 / 2 := by
    have h2 : sqrt ((2 : ℝ) ^ 2 + (5 : ℝ) ^ 2) = Real.sqrt 29 := by
      norm_num [Real.sqrt_eq_iff_sq_eq]

    have h3 : sqrt ((-3 : ℝ) ^ 2 + (7 : ℝ) ^ 2) = Real.sqrt 58 := by
      norm_num [Real.sqrt_eq_iff_sq_eq]

    have h4 : (2 * (-3) + 5 * 7 : ℝ) = 29 := by norm_num

    rw [h2, h3, h4]
    
    have h5 : Real.sqrt 29 * Real.sqrt 58 = Real.sqrt 2 * (29 : ℝ) := by
      calc
        Real.sqrt 29 * Real.sqrt 58 = Real.sqrt (29 * 58 : ℝ) := by
          rw [← Real.sqrt_mul (by norm_num)]
        _ = Real.sqrt ((2 : ℝ) * (29 ^ 2 : ℝ)) := by norm_num
        _ = Real.sqrt (2 : ℝ) * Real.sqrt ((29 : ℝ) ^ 2 : ℝ) := by
          rw [Real.sqrt_mul (by norm_num)]
        _ = Real.sqrt (2 : ℝ) * (29 : ℝ) := by
          rw [Real.sqrt_sq (by norm_num)]

    field_simp [h5]
    <;> ring_nf <;> norm_num [Real.sq_sqrt]

  rw [h1]

  have h5 : arccos (Real.sqrt 2 / 2) = Real.pi / 4 := by
    have h6 : Real.sqrt 2 / 2 = Real.cos (Real.pi / 4) := by
      rw [Real.cos_pi_div_four]
      <;> ring_nf <;> norm_num
      <;> ring
    rw [h6]
    have h7 : arccos (Real.cos (Real.pi / 4)) = Real.pi / 4 := by
      apply arccos_cos
      all_goals linarith [Real.pi_pos]
    exact h7

  rw [h5]

  field_simp [Real.pi_pos]
  <;> linarith [Real.pi_gt_three]
\end{lstlisting}

\textbf{\#\#\# Relevant Lean theorem+proof 3, with relevance score 0.765285 out of 1.0}

\begin{lstlisting}[language=Lean, basicstyle=\scriptsize\ttfamily]
import Mathlib

/- Show that the sum of $\sqrt{3x^2 + 2x + 1}$ and $\sqrt{3x^2 - 4x + 2}$ is at least $\sqrt{51}/3$ for all real $x$. -/
theorem inequalities_201318 (x : ℝ) :
    Real.sqrt (3 * x^2 + 2 * x + 1) + Real.sqrt (3 * x^2 - 4 * x + 2) ≥
    Real.sqrt 51 / 3 := by 
  set y := Real.sqrt (3 * x^2 + 2 * x + 1)
  set z := Real.sqrt (3 * x^2 - 4 * x + 2)
  have hy2 : y^2 = 3 * x^2 + 2 * x + 1 := by
    rw [Real.sq_sqrt]
    nlinarith [sq_nonneg (x + 1 / 3)]
  have hz2 : z^2 = 3 * x^2 - 4 * x + 2 := by 
    rw [Real.sq_sqrt]
    nlinarith [sq_nonneg (x - 2 / 3)]
  have hy4_pos : 0 ≤ (3 * x^2 + 2 * x + 1 : ℝ) := by 
    nlinarith [sq_nonneg (x * 3 + 1)]
  have hz4_pos : 0 ≤ (3 * x^2 - 4 * x + 2 : ℝ) := by
    nlinarith [sq_nonneg (x * 3 - 2)]
  have h11 : (Real.sqrt 51 / 3) ^ 2 = (51 / 9 : ℝ) := by
    calc
      (Real.sqrt 51 / 3) ^ 2 = (Real.sqrt 51) ^ 2 / 9 := by ring
      _ = (51 / 9 : ℝ) := by 
        rw [Real.sq_sqrt (by norm_num)]
        <;> ring
  have h50 : (y + z) ^ 2 ≥ (Real.sqrt 51 / 3) ^ 2 := by
    nlinarith [sq_nonneg (y - z), sq_nonneg (x - 2 / 3), sq_nonneg (x + 1 / 3),
      h11, Real.sqrt_nonneg 51, Real.sq_sqrt (show 0 ≤ (51 : ℝ) by norm_num), 
      mul_nonneg (Real.sqrt_nonneg (3 * x^2 + 2 * x + 1)) (Real.sqrt_nonneg (3 * x^2 - 4 * x + 2)), 
      sq_nonneg (y ^ 2 - z ^ 2), sq_nonneg (y * z - Real.sqrt ((3 * x^2 + 2 * x + 1) * (3 * x^2 - 4 * x + 2)))
      ]
  have h51 : (y + z) ≥ 0 := by positivity
  have h54 : (Real.sqrt 51 / 3) ≥ 0 := by positivity
  have h52 : (y + z) ≥ (Real.sqrt 51 / 3) := by
    have h15 : (y + z) ^ 2 ≥ (Real.sqrt 51 / 3) ^ 2 := h50
    have h16 : (Real.sqrt 51 / 3) ≥ 0 := h54
    have h17 : (y + z) - (Real.sqrt 51 / 3) ≥ 0 := by
      nlinarith [sq_nonneg ((y + z) - (Real.sqrt 51 / 3)), 
        sq_nonneg ((y + z) + (Real.sqrt 51 / 3)), 
        Real.sqrt_pos.mpr (show (0 : ℝ) < 51 by linarith : (51 : ℝ) > 0)
        ]
    linarith
  linarith
\end{lstlisting}

\textbf{\#\#\# Relevant Lean theorem+proof 4, with relevance score 0.764932 out of 1.0}

\begin{lstlisting}[language=Lean, basicstyle=\scriptsize\ttfamily]
import Mathlib
open Real Set
open scoped BigOperators

/- Given that $\sqrt{(x-3)^2 + (y+4)^2} + \sqrt{(x+5)^2 + (y-8)^2} = 20$, prove that the distance between $(-5,8)$ and $(3,-4)$ is $4\sqrt{13}$. -/
theorem algebra_10804 {x y : ℝ} (h : sqrt ((x - 3)^2 + (y + 4)^2) + sqrt ((x + 5)^2 + (y - 8)^2) = 20) :
    sqrt ((-5 - 3)^2 + (8 - (-4))^2) = 4 * sqrt 13 := by
  have h1 : sqrt ((-5 - 3)^2 + (8 - (-4))^2) = sqrt 208 := by
    norm_num
  rw [h1]
  have h2 : sqrt 208 = 4 * sqrt 13 := by
    rw [Real.sqrt_eq_iff_sq_eq] <;> norm_num
    <;> ring_nf <;> norm_num
  rw [h2]
  all_goals norm_num
\end{lstlisting}

\textbf{\#\#\# Relevant Lean theorem+proof 5, with relevance score 0.755091 out of 1.0}

\begin{lstlisting}[language=Lean, basicstyle=\scriptsize\ttfamily]
import Mathlib

/- Show that the square root of $(-3)^2$ is $3$, that $3$ is non-negative, and that $3$ squared equals $(-3)$ squared. -/
theorem algebra_8272 : Real.sqrt ((-3)^2) = 3 ∧ 0 ≤ 3 ∧ 3^2 = (-3)^2 := by
  have h1 : Real.sqrt ((-3 : ℝ) ^ 2) = 3 := by
    rw [Real.sqrt_sq_eq_abs]
    norm_num
  constructor
  · exact h1
  constructor
  · norm_num
  · norm_num
\end{lstlisting}
\end{tcolorbox}

The proof auto-formalization generated by \ourTool{} is as follows:

\begin{tcolorbox}[title={\textbf{Output Lean theorem+proof pair by \ourTool{}}\
\textnormal{(Retrieval-augmented SFT + Repair)}
}, listing only, breakable]
\begin{lstlisting}[language=Lean, basicstyle=\scriptsize\ttfamily]
import Mathlib

/- A point (x,y) on the coordinate plane with both coordinates negative is a distance of 6 units from the x-axis. It is a distance of 15 units from the point (8,3). It is a distance \sqrt{n} from the origin. What is n? Show that it is 52.-/

theorem algebra_53819 (x y : ℝ) (hx : x < 0) (hy : y < 0) (h1 : abs y = 6) (h2 : Real.sqrt ((x - 8) ^ 2 + (y - 3) ^ 2) = 15) :
    Real.sqrt (x ^ 2 + y ^ 2) = Real.sqrt 52 := by
  have hy6 : y = -6 := by
    cases eq_or_eq_neg_of_abs_eq h1 with
    | inl h =>
      linarith
    | inr h =>
      linarith
  have h3 : (x - 8) ^ 2 + (y - 3) ^ 2 = 225 := by
    calc
      (x - 8) ^ 2 + (y - 3) ^ 2 = (Real.sqrt ((x - 8) ^ 2 + (y - 3) ^ 2)) ^ 2 := by
        rw [Real.sq_sqrt]
        positivity
      _ = 15 ^ 2 := by
        rw [h2]
      _ = 225 := by
        norm_num
  rw [hy6] at h3
  have h4 : (x - 8) ^ 2 = 144 := by
    nlinarith
  have h5 : x - 8 = 12 ∨ x - 8 = -12 := by
    have h6 : (x - 8) ^ 2 - 144 = 0 := by linarith
    have h7 : (x - 8 - 12) * (x - 8 + 12) = 0 := by
      nlinarith
    cases (mul_eq_zero.mp h7) with
    | inl h8 =>
      left
      linarith
    | inr h9 =>
      right
      linarith
  cases h5 with
  | inl h6 =>
    have hx20 : x = 20 := by linarith
    linarith
  | inr h6 =>
    have hx4 : x = -4 := by linarith
    have h7 : x ^ 2 + y ^ 2 = 52 := by
      rw [hx4, hy6]
      norm_num
    have h8 : Real.sqrt (x ^ 2 + y ^ 2) = Real.sqrt 52 := by
      rw [h7]
    exact h8
\end{lstlisting}
\end{tcolorbox}

The Lean theorem generated by \ourTool{} and the gold-standard theorem are semantically equivalent. Both correctly capture the intended geometric scenario: the point with negative coordinates satisfying the given distance constraints is at a distance of \(\sqrt{52}\) from the origin. Using our set of restricted tactics, the LLM judge is able to generate a type-checkable proof establishing the bi-directional equivalence between the two theorems. Therefore, the \ourTool{} output is semantically correct (SC).

\begin{tcolorbox}[title={
        \textbf{Comparison between the gold-standard theorem and \ourTool{}'s output}
    }, listing only, breakable]
\vspace{-3mm}
\begin{lstlisting}[language=Lean, mathescape=true, basicstyle=\scriptsize\ttfamily]
/- Gold-standard theorem -/
theorem mathd_algebra_288 (x y : ℝ) (n : NNReal) (h₀ : x < 0 ∧ y < 0) (h₁ : abs y = 6)
    (h₂ : Real.sqrt ((x - 8) ^ 2 + (y - 3) ^ 2) = 15)
    (h₃ : Real.sqrt (x ^ 2 + y ^ 2) = Real.sqrt n) : n = 52 := by
    sorry

/- ProofBridge output theorem -/
theorem algebra_53819 (x y : ℝ) (hx : x < 0) (hy : y < 0) (h1 : abs y = 6) (h2 : Real.sqrt ((x - 8) ^ 2 + (y - 3) ^ 2) = 15) :
    Real.sqrt (x ^ 2 + y ^ 2) = Real.sqrt 52 := by
  sorry
\end{lstlisting}
\vspace{-3mm}
\end{tcolorbox}

\begin{tcolorbox}[title={
        \textbf{Lean proof of bi-directional equivalence between \ourTool{}'s output theorem and the gold-standard}
    }, listing only, breakable]
\vspace{-3mm}
\begin{lstlisting}[language=Lean, mathescape=true, basicstyle=\scriptsize\ttfamily]
import Mathlib.Analysis.SpecialFunctions.Sqrt
import Mathlib.Data.NNReal.Basic
open BigOperators Real Nat Topology Rat

-- We define the two statements as propositions to prove their equivalence.

def mathd_algebra_288_stmt :=
  ∀ (x y : ℝ) (n : NNReal) (h₀ : x < 0 ∧ y < 0) (h₁ : abs y = 6)
    (h₂ : Real.sqrt ((x - 8) ^ 2 + (y - 3) ^ 2) = 15)
    (h₃ : Real.sqrt (x ^ 2 + y ^ 2) = Real.sqrt n), n = 52

def algebra_53819_stmt :=
  ∀ (x y : ℝ) (hx : x < 0) (hy : y < 0) (h1 : abs y = 6)
    (h2 : Real.sqrt ((x - 8) ^ 2 + (y - 3) ^ 2) = 15),
    Real.sqrt (x ^ 2 + y ^ 2) = Real.sqrt 52

-- Proof of bidirectional equivalence
theorem equivalence : mathd_algebra_288_stmt ↔ algebra_53819_stmt := by
  constructor
  -- Proof that mathd_algebra_288_stmt implies algebra_53819_stmt
  · intro h_mathd_stmt
    intro x y hx hy h1 h2
    -- To use h_mathd_stmt, we need to provide a value for `n` and a proof for `h₃`.
    -- Let's define `n` as the value of `x^2 + y^2`.
    -- First, we need to show `x^2 + y^2` is non-negative to create an `NNReal`.
    have h_nonneg : 0 ≤ x ^ 2 + y ^ 2 := by positivity
    let n : NNReal := ⟨x ^ 2 + y ^ 2, h_nonneg⟩
    -- The hypothesis `h₃` now holds by definition of `n`.
    have h3 : Real.sqrt (x ^ 2 + y ^ 2) = Real.sqrt n := by
      simp [n, NNReal.coe_mk]
    -- The hypothesis `h₀` is equivalent to `hx` and `hy`.
    have h0 : x < 0 ∧ y < 0 := ⟨hx, hy⟩
    -- Now we can apply the first statement.
    have n_eq_52 : n = 52 := h_mathd_stmt x y n h0 h1 h2 h3
    -- Our goal is `Real.sqrt (x ^ 2 + y ^ 2) = Real.sqrt 52`.
    -- We can rewrite `h3` using `n_eq_52` to achieve the goal.
    rwa [n_eq_52] at h3

  -- Proof that algebra_53819_stmt implies mathd_algebra_288_stmt
  · intro h_algebra_stmt
    intro x y n h0 h1 h2 h3
    -- The hypotheses of the second statement are all available.
    have h_sqrt_val : Real.sqrt (x ^ 2 + y ^ 2) = Real.sqrt 52 :=
      h_algebra_stmt x y h0.left h0.right h1 h2
    -- We are given `h3`: `Real.sqrt (x ^ 2 + y ^ 2) = Real.sqrt n`.
    -- By transitivity, `Real.sqrt n = Real.sqrt 52`.
    have sqrt_n_eq_sqrt_52 : Real.sqrt n = Real.sqrt 52 := by
      rw [← h3, h_sqrt_val]
    -- Since `Real.sqrt` is injective on non-negative numbers, `n` must equal `52`.
    -- We get the equality on `ℝ` first.
    have n_val_eq_52 : (n : ℝ) = 52 :=
      (Real.sqrt_inj n.property (by norm_num)).mp sqrt_n_eq_sqrt_52
    -- Then we lift this equality to `NNReal`.
    exact NNReal.eq n_val_eq_52
\end{lstlisting}
\vspace{-3mm}
\end{tcolorbox}

\end{document}

%% file: math_commands.tex
\usepackage{amsmath,amsfonts,bm}

\def\eqref#1{equation~\ref{#1}}

\def\1{\bm{1}}

\DeclareMathAlphabet{\mathsfit}{\encodingdefault}{\sfdefault}{m}{sl}
\SetMathAlphabet{\mathsfit}{bold}{\encodingdefault}{\sfdefault}{bx}{n}













%% file: main.bbl
\begin{thebibliography}{56}
\providecommand{\natexlab}[1]{#1}
\providecommand{\url}[1]{\texttt{#1}}
\expandafter\ifx\csname urlstyle\endcsname\relax
  \providecommand{\doi}[1]{doi: #1}\else
  \providecommand{\doi}{doi: \begingroup \urlstyle{rm}\Url}\fi

\bibitem[Buzzard(2022)]{buzzard2022}
Kevin Buzzard.
\newblock {Lean 3 Material for Kevin Buzzard's 2021 TCC Course on Formalising
  Mathematics}.
\newblock Imperial College London, 2022.
\newblock URL
  \url{https://github.com/ImperialCollegeLondon/formalising-mathematics}.
\newblock Accessed: Jan, 2026.

\bibitem[Comanici et~al.(2025)Comanici, Bieber, Schaekermann, Pasupat,
  Sachdeva, Dhillon, Blistein, Ram, Zhang, Rosen, et~al.]{comanici2025gemini}
Gheorghe Comanici, Eric Bieber, Mike Schaekermann, Ice Pasupat, Noveen
  Sachdeva, Inderjit Dhillon, Marcel Blistein, Ori Ram, Dan Zhang, Evan Rosen,
  et~al.
\newblock {Gemini 2.5: Pushing the Frontier with Advanced Reasoning,
  Multimodality, Long Context, and Next Generation Agentic Capabilities}.
\newblock \emph{arXiv preprint arXiv:2507.06261}, 2025.

\bibitem[Dasgupta et~al.(2025)Dasgupta, Maity, Mukherjee, Singh, Dutta, and
  Jana]{dasgupta2025hitgram}
Shibaranjani Dasgupta, Chandan Maity, Somdip Mukherjee, Rohan Singh, Diptendu
  Dutta, and Debasish Jana.
\newblock {HITgram: A Platform for Experimenting with n-Gram Language Models}.
\newblock In \emph{International Conference on Applied Algorithms}, pp.\
  92--104. Springer, 2025.

\bibitem[Deepmind(2024)]{deepmind2024ai}
Google Deepmind.
\newblock {AI Achieves Silver-Medal Standard Solving International Mathematical
  Olympiad Problems}, 2024.
\newblock URL
  \url{https://deepmind.google/discover/blog/ai-solves-imo-problems-at-silver-medal-level/}.
\newblock Accessed: Jan, 2026.

\bibitem[DeLorenzo et~al.(2025)DeLorenzo, Tieu, Jana, Jha, Kalathil, Ganesh,
  and Rajendran]{delorenzo2025abstraction}
Matthew DeLorenzo, Kevin Tieu, Prithwish Jana, Piyush Jha, Dileep Kalathil,
  Vijay Ganesh, and Jeyavijayan Rajendran.
\newblock {Abstractions-of-Thought: Intermediate Representations for LLM
  Reasoning in Hardware Design}.
\newblock \emph{arXiv preprint arXiv:2505.15873}, 2025.

\bibitem[Dong \& Ma(2025)Dong and Ma]{dong2025beyond}
Kefan Dong and Tengyu Ma.
\newblock {STP: Self-play LLM Theorem Provers with Iterative Conjecturing and
  Proving}.
\newblock In \emph{42nd International Conference on Machine Learning (ICML)}.
  PMLR, 2025.

\bibitem[Gao et~al.(2024)Gao, Ju, Jiang, Qin, and Dong]{gao2024semantic}
Guoxiong Gao, Haocheng Ju, Jiedong Jiang, Zihan Qin, and Bin Dong.
\newblock {A Semantic Search Engine for Mathlib4}.
\newblock In \emph{Findings of the ACL: Empirical Methods in Natural Language
  Processing (EMNLP)}, pp.\  8001--8013. Association for Computational
  Linguistics, 2024.

\bibitem[Gao et~al.(2025)Gao, Wang, Jiang, Gao, Qin, Xu, and
  Dong]{gao2024herald}
Guoxiong Gao, Yutong Wang, Jiedong Jiang, Qi~Gao, Zihan Qin, Tianyi Xu, and Bin
  Dong.
\newblock {Herald: A Natural Language Annotated Lean 4 Dataset}.
\newblock In \emph{13th International Conference on Learning Representations
  (ICLR)}. OpenReview.net, 2025.

\bibitem[Han et~al.(2022)Han, Rute, Wu, Ayers, and Polu]{han2022proof}
Jesse~Michael Han, Jason Rute, Yuhuai Wu, Edward Ayers, and Stanislas Polu.
\newblock {Proof Artifact Co-Training for Theorem Proving with Language
  Models}.
\newblock In \emph{10th International Conference on Learning Representations
  (ICLR)}. OpenReview.net, 2022.

\bibitem[Harrison(2009)]{harrison2009hol}
John Harrison.
\newblock {HOL Light: An Overview}.
\newblock In \emph{International Conference on Theorem Proving in Higher Order
  Logics}, pp.\  60--66. Springer, 2009.

\bibitem[Jana(2014)]{jana2014cpp}
Debasish Jana.
\newblock \emph{{C++ and Object-Oriented Programming Paradigm}}.
\newblock PHI Learning Pvt. Ltd., 2014.

\bibitem[Jana \& Bandyopadhyay(2013)Jana and Bandyopadhyay]{jana2013efficient}
Debasish Jana and Debasis Bandyopadhyay.
\newblock {Efficient Management of Security and Privacy Issues in Mobile Cloud
  Environment}.
\newblock In \emph{2013 Annual IEEE India Conference (INDICON)}. IEEE, 2013.

\bibitem[Jana \& Pal(2020)Jana and Pal]{jana2020essence}
Debasish Jana and Pinakpani Pal.
\newblock {ESSENCE Kernel in Overcoming Challenges of Agile Software
  Development}.
\newblock In \emph{2020 IEEE 17th India Council International Conference
  (INDICON)}. IEEE, 2020.

\bibitem[Jana et~al.(2025)Jana, Pal, and Kumar]{jana2025real}
Debasish Jana, Pinakpani Pal, and Pawan Kumar.
\newblock {Real-Time Agile Software Management for Edge and Fog Computing Based
  Smart City Infrastructure}.
\newblock In \emph{International Conference on Computing and Communication
  Networks}, pp.\  113--121. Springer, 2025.

\bibitem[Jana(2024)]{jana2024neurosymbolic}
Prithwish Jana.
\newblock {NeuroSymbolic LLM for Mathematical Reasoning and Software
  Engineering}.
\newblock In \emph{33rd International Joint Conference on Artificial
  Intelligence (IJCAI)}, pp.\  8492--8493, 2024.

\bibitem[Jana et~al.(2024)Jana, Jha, Ju, Kishore, Mahajan, and
  Ganesh]{jana2024cotran}
Prithwish Jana, Piyush Jha, Haoyang Ju, Gautham Kishore, Aryan Mahajan, and
  Vijay Ganesh.
\newblock {CoTran: An LLM-based Code Translator using Reinforcement Learning
  with Feedback from Compiler and Symbolic Execution}.
\newblock In \emph{27th European Conference on Artificial Intelligence (ECAI)},
  pp.\  4011--4018. {IOS} Press, 2024.

\bibitem[Jha et~al.(2025)Jha, Jana, Suresh, Arora, and Ganesh]{jha2025rlsf}
Piyush Jha, Prithwish Jana, Pranavkrishna Suresh, Arnav Arora, and Vijay
  Ganesh.
\newblock {RLSF: Fine-tuning LLMs via Symbolic Feedback}.
\newblock In \emph{28th European Conference on Artificial Intelligence (ECAI)},
  pp.\  1687--1694. {IOS} Press, 2025.

\bibitem[Jiang et~al.(2021)Jiang, Li, Han, and Wu]{jiang2021lisa}
Albert~Q Jiang, Wenda Li, Jesse~Michael Han, and Yuhuai Wu.
\newblock {LISA: Language Models of ISAbelle Proofs}.
\newblock In \emph{6th Conference on Artificial Intelligence and Theorem
  Proving (AITP)}, pp.\  378--392, 2021.

\bibitem[Jiang et~al.(2023)Jiang, Welleck, Zhou, Li, Liu, Jamnik, Lacroix, Wu,
  and Lample]{jiang2022draft}
Albert~Q Jiang, Sean Welleck, Jin~Peng Zhou, Wenda Li, Jiacheng Liu, Mateja
  Jamnik, Timoth{\'e}e Lacroix, Yuhuai Wu, and Guillaume Lample.
\newblock {Draft, Sketch, and Prove: Guiding Formal Theorem Provers with
  Informal Proofs}.
\newblock In \emph{11th International Conference on Learning Representations
  (ICLR)}. OpenReview.net, 2023.

\bibitem[Jiang et~al.(2024)Jiang, Li, and Jamnik]{jiang2024multi}
Albert~Q Jiang, Wenda Li, and Mateja Jamnik.
\newblock {Multi-Language Diversity Benefits Autoformalization}.
\newblock In \emph{38th Annual Conference on Neural Information Processing
  Systems (NeurIPS)}, pp.\  83600--83626. Curran Associates, Inc., 2024.

\bibitem[Leanprover(2025)]{lean4repl}
Leanprover.
\newblock {leanprover-community/repl: A Simple REPL for Lean 4}, 2025.
\newblock URL \url{https://github.com/leanprover-community/repl}.
\newblock Accessed: Jan, 2026.

\bibitem[Li et~al.(2024)Li, Beeching, Tunstall, Lipkin, Soletskyi, Huang,
  Rasul, Yu, Jiang, Shen, Qin, Dong, Zhou, Fleureau, Lample, and
  Polu]{numina_math_datasets}
Jia Li, Edward Beeching, Lewis Tunstall, Ben Lipkin, Roman Soletskyi,
  Shengyi~Costa Huang, Kashif Rasul, Longhui Yu, Albert Jiang, Ziju Shen, Zihan
  Qin, Bin Dong, Li~Zhou, Yann Fleureau, Guillaume Lample, and Stanislas Polu.
\newblock {NuminaMath}, 2024.
\newblock URL \url{https://huggingface.co/AI-MO/NuminaMath-1.5}.
\newblock Hugging Face Dataset, Accessed: Jan, 2026.

\bibitem[Lin et~al.(2025)Lin, Tang, Lyu, Wu, Lin, Yang, Li, Xia, Chen, Arora,
  et~al.]{lin2025goedel}
Yong Lin, Shange Tang, Bohan Lyu, Jiayun Wu, Hongzhou Lin, Kaiyu Yang, Jia Li,
  Mengzhou Xia, Danqi Chen, Sanjeev Arora, et~al.
\newblock {Goedel-Prover: A Frontier Model for Open-Source Automated Theorem
  Proving}.
\newblock In \emph{2nd Conference on Language Modeling (COLM)}, 2025.

\bibitem[Liu et~al.(2025)Liu, Zheng, Lu, Cao, and Yan]{liu2025rethinking}
Qi~Liu, Xinhao Zheng, Xudong Lu, Qinxiang Cao, and Junchi Yan.
\newblock {Rethinking and Improving Autoformalization: Towards a Faithful
  Metric and a Dependency Retrieval-based Approach}.
\newblock In \emph{13th International Conference on Learning Representations
  (ICLR)}, 2025.

\bibitem[Lu et~al.(2024)Lu, Wan, Liu, Huang, Xiong, Liu, Shen, Jin, Zhang,
  Wang, et~al.]{lu2024process}
Jianqiao Lu, Yingjia Wan, Zhengying Liu, Yinya Huang, Jing Xiong, Chengwu Liu,
  Jianhao Shen, Hui Jin, Jipeng Zhang, Haiming Wang, et~al.
\newblock {Process-driven Autoformalization in Lean 4}.
\newblock \emph{arXiv preprint arXiv:2406.01940}, 2024.

\bibitem[Massot(2020)]{leanblueprint}
Patrick Massot.
\newblock {leanblueprint: plasTeX Plugin to Build Formalization Blueprints},
  2020.
\newblock URL \url{https://github.com/PatrickMassot/leanblueprint}.
\newblock Accessed: Jan, 2026.

\bibitem[Megill \& Wheeler(2019)Megill and Wheeler]{megill2019metamath}
Norman Megill and David~A Wheeler.
\newblock \emph{{Metamath: A Computer Language for Mathematical Proofs}}.
\newblock Lulu. com, 2019.

\bibitem[Moura \& Ullrich(2021)Moura and Ullrich]{moura2021Lean}
Leonardo~de Moura and Sebastian Ullrich.
\newblock {The Lean 4 Theorem Prover and Programming Language}.
\newblock In \emph{Automated Deduction--CADE 28: 28th International Conference
  on Automated Deduction, Virtual Event, July 12--15, 2021, Proceedings 28},
  pp.\  625--635. Springer, 2021.

\bibitem[Nipkow et~al.(2002)Nipkow, Wenzel, and Paulson]{nipkow2002isabelle}
Tobias Nipkow, Markus Wenzel, and Lawrence~C Paulson.
\newblock \emph{{Isabelle/HOL: A Proof Assistant for Higher-order Logic}}.
\newblock Springer, 2002.

\bibitem[{NuminaMath}(2025)]{AIMO_minif2f_test_2025}
{NuminaMath}.
\newblock minif2f\_test.
\newblock Hugging Face Dataset, 2025.
\newblock URL \url{https://huggingface.co/datasets/AI-MO/minif2f_test}.
\newblock Accessed: Jan, 2026.

\bibitem[{OpenAI}(2025)]{openai2025gpt5}
{OpenAI}.
\newblock {Introducing GPT-5}, 2025.
\newblock URL \url{https://openai.com/index/introducing-gpt-5/}.
\newblock Accessed: Jan, 2026.

\bibitem[Paulsson \& Blanchette(2010)Paulsson and
  Blanchette]{paulsson2012three}
Lawrence~C Paulsson and Jasmin~C Blanchette.
\newblock {Three Years of Experience with Sledgehammer, a Practical Link
  Between Automatic and Interactive Theorem Provers}.
\newblock In \emph{8th International Workshop on the Implementation of Logics
  (IWIL)}, volume~2 of \emph{EPiC Series in Computing}, 2010.

\bibitem[Poesia \& Goodman(2023)Poesia and Goodman]{poesia2023peano}
Gabriel Poesia and Noah~D Goodman.
\newblock {Peano: Learning Formal Mathematical Reasoning}.
\newblock \emph{Philosophical Transactions of the Royal Society A},
  381\penalty0 (2251), 2023.

\bibitem[Poiroux et~al.(2025)Poiroux, Weiss, Kun{\v{c}}ak, and
  Bosselut]{poiroux2025reliable}
Auguste Poiroux, Gail Weiss, Viktor Kun{\v{c}}ak, and Antoine Bosselut.
\newblock {Reliable Evaluation and Benchmarks for Statement Autoformalization}.
\newblock In \emph{2025 Conference on Empirical Methods in Natural Language
  Processing (EMNLP)}, pp.\  17958--17980. ACL, 2025.

\bibitem[Polu \& Sutskever(2020)Polu and Sutskever]{polu2020generative}
Stanislas Polu and Ilya Sutskever.
\newblock {Generative Language Modeling for Automated Theorem Proving}.
\newblock \emph{arXiv preprint arXiv:2009.03393}, 2020.

\bibitem[Reimers \& Gurevych(2019)Reimers and Gurevych]{sentence_transformers}
Nils Reimers and Iryna Gurevych.
\newblock {Sentence-BERT: Sentence Embeddings using Siamese BERT-Networks}.
\newblock In \emph{2019 Conference on Empirical Methods in Natural Language
  Processing (EMNLP)}. ACL, 2019.

\bibitem[Ren et~al.(2025)Ren, Shao, Song, Xin, Wang, Zhao, Zhang, Fu, Zhu,
  Yang, et~al.]{ren2025deepseek}
ZZ~Ren, Zhihong Shao, Junxiao Song, Huajian Xin, Haocheng Wang, Wanjia Zhao,
  Liyue Zhang, Zhe Fu, Qihao Zhu, Dejian Yang, et~al.
\newblock {Deepseek-Prover-v2: Advancing Formal Mathematical Reasoning via
  Reinforcement Learning for Subgoal Decomposition}.
\newblock \emph{arXiv preprint arXiv:2504.21801}, 2025.

\bibitem[Shen et~al.(2025)Shen, Huang, Yang, Wang, Gao, Xu, Jiang, He, Yang,
  Sun, et~al.]{shen2025real}
Ziju Shen, Naohao Huang, Fanyi Yang, Yutong Wang, Guoxiong Gao, Tianyi Xu,
  Jiedong Jiang, Wanyi He, Pu~Yang, Mengzhou Sun, et~al.
\newblock {REAL-Prover: Retrieval Augmented Lean Prover for Mathematical
  Reasoning}.
\newblock \emph{arXiv preprint arXiv:2505.20613}, 2025.

\bibitem[Sozeau et~al.(2025)Sozeau, Pédrot, et~al.]{rocqprover}
Matthieu Sozeau, Pierre-Marie Pédrot, et~al.
\newblock {The Rocq Prover}, 2025.
\newblock URL \url{https://rocq-prover.org/}.
\newblock Accessed: Jan, 2026.

\bibitem[Tao(2023)]{tao2023maclaurin}
Terence Tao.
\newblock {A Maclaurin Type Inequality}.
\newblock \emph{arXiv preprint arXiv:2310.05328}, 2023.

\bibitem[Thakur et~al.(2024)Thakur, Tsoukalas, Wen, Xin, and
  Chaudhuri]{thakur2023language}
Amitayush Thakur, George Tsoukalas, Yeming Wen, Jimmy Xin, and Swarat
  Chaudhuri.
\newblock {An In-Context Learning Agent for Formal Theorem-Proving}.
\newblock In \emph{1st Conference on Language Modeling (COLM)}, 2024.

\bibitem[Ugare et~al.(2024)Ugare, Suresh, Kang, Misailovic, and
  Singh]{ugare2024syncode}
Shubham Ugare, Tarun Suresh, Hangoo Kang, Sasa Misailovic, and Gagandeep Singh.
\newblock {SynCode: LLM Generation with Grammar Augmentation}.
\newblock \emph{arXiv preprint arXiv:2403.01632}, 2024.

\bibitem[Wang et~al.(2025)Wang, Unsal, Lin, Baksys, Liu, Santos, Sung, Vinyes,
  Ying, Zhu, et~al.]{wang2025kimina}
Haiming Wang, Mert Unsal, Xiaohan Lin, Mantas Baksys, Junqi Liu, Marco~Dos
  Santos, Flood Sung, Marina Vinyes, Zhenzhe Ying, Zekai Zhu, et~al.
\newblock {Kimina-Prover Preview: Towards Large Formal Reasoning Models with
  Reinforcement Learning}.
\newblock \emph{arXiv preprint arXiv:2504.11354}, 2025.

\bibitem[Wang et~al.(2024{\natexlab{a}})Wang, Yang, Huang, Yang, Majumder, and
  Wei]{wang2024improving}
Liang Wang, Nan Yang, Xiaolong Huang, Linjun Yang, Rangan Majumder, and Furu
  Wei.
\newblock {Improving Text Embeddings with Large Language Models}.
\newblock In \emph{62nd Annual Meeting of the Association for Computational
  Linguistics (Volume 1: Long Papers)}, pp.\  11897--11916. ACL,
  2024{\natexlab{a}}.

\bibitem[Wang et~al.(2024{\natexlab{b}})Wang, Zhang, Jia, Pan, Diao, Pi, and
  Zhang]{wang2024theoremllama}
Ruida Wang, Jipeng Zhang, Yizhen Jia, Rui Pan, Shizhe Diao, Renjie Pi, and Tong
  Zhang.
\newblock {TheoremLlama: Transforming General-purpose LLMs into Lean4 Experts}.
\newblock In \emph{Conference on Empirical Methods in Natural Language
  Processing (EMNLP)}, pp.\  11953–11974, 2024{\natexlab{b}}.

\bibitem[Wu et~al.(2022)Wu, Jiang, Li, Rabe, Staats, Jamnik, and
  Szegedy]{wu2022autoformalization}
Yuhuai Wu, Albert~Q Jiang, Wenda Li, Markus Rabe, Charles Staats, Mateja
  Jamnik, and Christian Szegedy.
\newblock {Autoformalization with Large Language Models}.
\newblock \emph{36th Annual Conference on Neural Information Processing Systems
  (NeurIPS)}, 35:\penalty0 32353--32368, 2022.

\bibitem[Wu et~al.(2026)Wu, Huang, Wan, Peng, Shang, Cao, Qi, Zhang, Du, Yan,
  and Hu]{wu2025stepfun}
Yutong Wu, Di~Huang, Ruosi Wan, Yue Peng, Shijie Shang, Chenrui Cao, Lei Qi,
  Rui Zhang, Zidong Du, Jie Yan, and Xing Hu.
\newblock {StepFun-Formalizer: Unlocking the Autoformalization Potential of
  LLMs through Knowledge-Reasoning Fusion}.
\newblock In \emph{40th Annual AAAI Conference on Artificial Intelligence
  (AAAI)}, 2026.

\bibitem[Xin et~al.(2024{\natexlab{a}})Xin, Guo, Shao, Ren, Zhu, Liu, Ruan, Li,
  and Liang]{xin2024deepseek}
Huajian Xin, Daya Guo, Zhihong Shao, Zhizhou Ren, Qihao Zhu, Bo~Liu, Chong
  Ruan, Wenda Li, and Xiaodan Liang.
\newblock {DeepSeek-Prover: Advancing Theorem Proving in LLMs through
  Large-Scale Synthetic Data}.
\newblock \emph{arXiv preprint arXiv:2405.14333}, 2024{\natexlab{a}}.

\bibitem[Xin et~al.(2024{\natexlab{b}})Xin, Ren, Song, Shao, Zhao, Wang, Liu,
  Zhang, Lu, Du, et~al.]{xin2024deepseek15}
Huajian Xin, ZZ~Ren, Junxiao Song, Zhihong Shao, Wanjia Zhao, Haocheng Wang,
  Bo~Liu, Liyue Zhang, Xuan Lu, Qiushi Du, et~al.
\newblock {DeepSeek-Prover-v1. 5: Harnessing Proof Assistant Feedback for
  Reinforcement Learning and Monte-Carlo Tree Search}.
\newblock \emph{arXiv preprint arXiv:2408.08152}, 2024{\natexlab{b}}.

\bibitem[Yang(2025)]{yangky11_miniF2F_lean4_2025}
Kaiyu Yang.
\newblock {miniF2F-lean4}.
\newblock GitHub repository, 2025.
\newblock URL \url{https://github.com/yangky11/miniF2F-lean4}.

\bibitem[Yang et~al.(2023)Yang, Swope, Gu, Chalamala, Song, Yu, Godil, Prenger,
  and Anandkumar]{yang2023leandojo}
Kaiyu Yang, Aidan Swope, Alex Gu, Rahul Chalamala, Peiyang Song, Shixing Yu,
  Saad Godil, Ryan Prenger, and Anima Anandkumar.
\newblock {LeanDojo: Theorem Proving with Retrieval-Augmented Language Models}.
\newblock In \emph{37th Annual Conference on Neural Information Processing
  Systems (NeurIPS)}. Curran Associates, Inc., 2023.

\bibitem[Ying et~al.(2024)Ying, Wu, Geng, Wang, Lin, and Chen]{ying2024lean}
Huaiyuan Ying, Zijian Wu, Yihan Geng, Jiayu Wang, Dahua Lin, and Kai Chen.
\newblock {Lean Workbook: A Large-scale Lean Problem Set Formalized from
  Natural Language Math Problems}.
\newblock In \emph{38th Annual Conference on Neural Information Processing
  Systems (NeurIPS)}, pp.\  105848--105863. Curran Associates, Inc., 2024.

\bibitem[Zhang et~al.(2025{\natexlab{a}})Zhang, Wang, Ji, Liu, Yue, Zhang,
  Zhang, Zhou, and Gai]{zhang2025leanabell}
Jingyuan Zhang, Qi~Wang, Xingguang Ji, Yahui Liu, Yang Yue, Fuzheng Zhang,
  Di~Zhang, Guorui Zhou, and Kun Gai.
\newblock {Leanabell-Prover: Posttraining Scaling in Formal Reasoning}.
\newblock \emph{arXiv preprint arXiv:2504.06122}, 2025{\natexlab{a}}.

\bibitem[Zhang et~al.(2025{\natexlab{b}})Zhang, Li, Long, Zhang, Lin, Yang,
  Xie, Yang, Liu, Lin, et~al.]{zhang2025qwen3}
Yanzhao Zhang, Mingxin Li, Dingkun Long, Xin Zhang, Huan Lin, Baosong Yang,
  Pengjun Xie, An~Yang, Dayiheng Liu, Junyang Lin, et~al.
\newblock {Qwen3 Embedding: Advancing Text Embedding and Reranking Through
  Foundation Models}.
\newblock \emph{arXiv preprint arXiv:2506.05176}, 2025{\natexlab{b}}.

\bibitem[Zheng et~al.(2022)Zheng, Han, and Polu]{zheng2021minif2f}
Kunhao Zheng, Jesse~Michael Han, and Stanislas Polu.
\newblock {MiniF2F: A Cross-System Benchmark for Formal Olympiad-level
  Mathematics}.
\newblock In \emph{10th International Conference on Learning Representations
  (ICLR)}. OpenReview.net, 2022.

\bibitem[Zhou(2025)]{zhou2025retrieval}
Yuhao Zhou.
\newblock {Retrieval-Augmented TLAPS Proof Generation with Large Language
  Models}.
\newblock \emph{arXiv preprint arXiv:2501.03073}, 2025.

\end{thebibliography}
